\documentclass[%
 aip,
 jmp,%
 amsmath,amssymb,
preprint,%
]{revtex4-1}

\usepackage{graphicx}
\usepackage{dcolumn}
\usepackage{bm}
\usepackage{color}

\begin{document}


\title{Zero Prandtl-number rotating magnetoconvection}

\author{Manojit Ghosh}
\author{Pinaki Pal}%
 \email{pinaki.pal@maths.nitdgp.ac.in.}
\affiliation{Department of Mathematics, National Institute of Technology, Durgapur-713209, India}


\date{\today}

\begin{abstract}
We investigate instabilities and chaos near the onset of Rayleigh-B\'{e}nard convection (RBC) of electrically conducting fluids with free-slip, perfectly electrically and thermally conducting boundary conditions in presence of uniform rotation about vertical axis and horizontal external magnetic field by considering zero Prandtl-number  limit ($\mathrm{Pr} \rightarrow 0$). Direct numerical simulations (DNS) and low dimensional modeling of the system are done for the investigation.  Values of  the Chandrasekhar number ($\mathrm{Q}$) and the Taylor number ($\mathrm{Ta}$) are varied in the range $0 <  \mathrm{Q}, \mathrm{Ta} \leq 50$.  Depending on the values of the parameters in the chosen range and choice of initial conditions, onset of convection is found be either periodic or chaotic. Interestingly, it is found that chaos at the onset can occur through four different routes namely homoclinic, intermittency, period doubling and quasi-periodic routes. Homoclinic and intermittent routes to chaos at the onset occur in presence of weak magnetic field ($\mathrm{Q} < 2$), while period doubling route is observed for relatively stronger magnetic field ($\mathrm{Q} \geq 2$) for one set of initial conditions. On the other hand, quasiperiodic route to chaos at the onset is observed for another set of initial conditions. However, the rotation rate (value of $\mathrm{Ta}$) also plays an important role in determining the nature of convection at the onset. Analysis of the system simultaneously with DNS and low dimensional modeling helps to clearly identify different flow regimes concentrated near the onset of convection and understand their origins. The periodic or chaotic convection at the onset is found to be connected with rich bifurcation structures involving subcritical pitchfork, imperfect pitchfork,  supercritical Hopf, imperfect homoclinic gluing and Neimark-Sacker bifurcations.
%
\end{abstract}

\maketitle


\section{Introduction}
Convection is known to play an important role in the dynamics of many natural as well as technological systems. Some examples of convective system include oceans~\cite{marshall:1999}, atmosphere~\cite{hartman:2001}, geophysical~\cite{cardin:1994,glatzmaier:1995} and astrophysical~\cite{cattaneo:2003} systems besides the technological systems like metal production~\cite{brent:1988} and crystal growth~\cite{hurle:1994}. Understanding different convective phenomena in these systems is of utmost importance and researchers consider a simplified version of convection called Rayleigh-B\'{e}nard convection (RBC) for this purpose. RBC in its simplest form consists of a layer of fluid kept between two infinite, horizontal, conducting plates and heated from below~\cite{chandra:book,bodenschatz:2000,ahlers:2009}. Two dimensionless parameters namely the Rayleigh-number ($\mathrm{Ra}$, vigor of buoyancy) and the Prandtl-number ($\mathrm{Pr}$, ratio of thermal diffusion and viscous relaxation time scales) appear in the mathematical description of RBC. RBC provides a paradigmatic model to investigate several aspects of convection including instabilities, pattern formation, bifurcations, chaos, turbulence and many more for different ranges of values of $\mathrm{Ra}$ and $\mathrm{Pr}$~\cite{chandra:book,bodenschatz:2000,ahlers:2009,swinney_gollub:book_1985}. Over the years, extensive studies on RBC has not only improved the understanding of the basic physics of convection but also enriched the theory of hydrodynamic stability~\cite{chandra:book,drazin:book}, pattern formation~\cite{bodenschatz:2000,mannevile:book,cross_hohenberg:1993} and spatio temporal chaos~\cite{bodenschatz:2000,getling:book}. 

Chandrasekhar~\cite{chandra:book} theoretically investigated the problem of the onset of RBC with free-slip and rigid boundary conditions. Performing linear analysis, he found that at the onset of RBC, principle of exchange of stabilities is valid and stationary cellular convection is prevalent. He analytically determined the critical values of the Rayleigh number and wave number at the onset of RBC which are independent of the Prandtl number. However, stability of the patterns at the onset of convection can not be determined from the linear analysis. Later, Schl\"{u}ter et al.~\cite{schluter:1965} theoretically established through nonlinear analysis that stable pattern at the onset of convection is straight two dimensional (2D) rolls. The stability regime of these 2D-rolls above the onset of convection in the $\mathrm{Ra} - \mathrm{k}~(\mathrm{wavenumber})$-space as a function of $\mathrm{Pr}$ i.e. the so called `Busse balloon' has been determined by Busse and his collaborators~\cite{swinney_gollub:book_1985,busse:1978,busse:1989}. RBC experiments performed with silicon oil (large $\mathrm{Pr}$) and gases (small $\mathrm{Pr}$) are found to agree well with `Busse balloon'~\cite{busse:1971,croquette1:Contemp.Phys_1989,croquette2:Contemp.Phys_1989}. Numerous theoretical, numerical and experimental works performed with RBC model have investigated convection related issues including heat transfer, pattern formation and turbulence(for a detailed review please see~\cite{bodenschatz:2000,ahlers:2009,swinney_gollub:book_1985}). It is evident from the computation of `Busse balloon' that Prandtl number has important role in determining secondary and higher order instabilities. In fact, as the value of the Prandtl number decreases, secondary and higher order instabilities approach towards the primary one and a rich dynamics is expected very close to the onset of convection~\cite{Busse:1972,clever:POF_1990,thual:1992,kft,mishra:2010}. 
The theoretical as well as numerical investigations performed in~\cite{pal:2002,pal:2009,paul:2011,pal:2013,dan:2014,dan:2015,nandu:2016} for low Prandtl number fluids have revealed expected rich bifurcation structure within few percent above the onset of convection. 

Researchers also carried out investigations of convective phenomena using RBC model in presence of rotation or/and magnetic field in order to mimic geophysical and astrophysical situations more closely~\cite{chandra:book,spiegel:JGR_1962,rossby:JFM_1969,Roberts:1975}.  Mathematical description of RBC needs three more dimensionless parameters apart from $\mathrm{Ra}$ and $\mathrm{Pr}$ in this case, namely the Taylor number ($\mathrm{Ta}$, strength of the Coriolis force), the Chandrasekhar number  ($\mathrm{Q}$, strength of the Lorenz force) and the magnetic Prandtl-number ($\mathrm{Pm}$, ratio of magnetic diffusion time scale and viscous relaxation time scale). Interestingly, it has been discovered through theoretical~\cite{chandra:book} and experimental~\cite{nakagawa_a:1955,nakagawa:1955} investigations that both rotation and magnetic field, acting separately, inhibit the onset of instability and elongate the rolls which appear at marginal stability when the magnetic field is applied in the vertical direction~\cite{chandra:book}. Horizontal magnetic field in absence of rotation, on the other hand does not change the primary instability but significantly influence the higher order instabilities~\cite{chandra:book}. Theoretical~\cite{busse:1983} as well as experimental~\cite{fauve_prl:1984,fauve:1984,burr:2002,yanagisawa:2010} investigations performed with electrically conducting low Prandtl number fluids in presence of horizontal magnetic field revealed very rich dynamics including homoclinic bifurcations, period doubling cascade etc. near the onset of convection.

Chandrasekhar first investigated RBC in simultaneous presence of rotation and uniform external magnetic field and  determined critical Rayleigh number and wave number at the onset of convection for wide ranges of values of $\mathrm{Ta}$ and $\mathrm{Q}$ from linear theory. The theoretical predictions of Chandrasekhar have been experimentally verified by Nakagawa~\cite{nakagawa:1957,nakagawa:1959} for large values of Taylor number and Chandrasekhar number. Later, Eltayeb~\cite{Eltayeb:1972} performed extensive linear analysis of the same system in the asymptotic limits of $\mathrm{Q},\mathrm{Ta}\rightarrow \infty$ for various boundary conditions and orientations of the external magnetic field and rotation. He showed that $\mathrm{Q}$ and $\mathrm{Ta}$ follow some well defined power laws. Subsequently, Roberts and Stewartson~\cite{Roberts:1975} theoretically studied rotating magnetoconvection with Rayleigh-B\'{e}nard geometry in presence of rotation about vertical axis and horizontal uniform magnetic field using free-slip velocity boundary conditions. They reported the existence of oblique rolls and studied their stability. Soward~\cite{Soward:1980} theoretically investigated similar rotating magneto-convection (RMC) system with no-slip velocity boundary conditions and also reported the existence of stable oblique rolls in some regions of the parameter space. Aurnou and Olson~\cite{Olson:2001} studied RMC in liquid gallium ($\mathrm{Pr} = 0.023$) experimentally and found heat transfer is inhibited for high Taylor number.  Theoretical investigation of Zhang et al.~\cite{zhang:2004} agree well with the experimental findings of Aurnou and Olson. RMC has been numerically investigated for heat transfer properties of low Prandtl number electrically conducting fluids both with horizontal and vertical magnetic field by Varshney and Baig~\cite{Baig:2008,Baig1:2008}. Podgivina~\cite{Podvigina:2008,Podvigina:2010} studied the stability of rolls in RMC of electrically conducting fluids  with an imposed vertical magnetic field in a rigid electrically insulating horizontal boundaries. She determined regions of the parameter space where rolls emerge at the onset of convection. Recently,  Eltayeb and Rahaman~\cite{Eltayeb:2013}, studied RMC in the presence of horizontal magnetic field and rotation using linear theory for various boundary conditions and determined preferred mode of convection in each case. 

The effects of rotation and external magnetic field on the instabilities and bifurcation structures near the onset of convection when acted separately have recently been investigated in~\cite{pal:2012,hirdesh:2013,priyanka:2013,priyanka:2014,arnab:2014,nandu:2015,arnab:2015,arnab:2016} and reported very rich dynamics including chaos at the onset of convection. However, instabilities and the associated bifurcation structures near the onset of RBC of electrically conducting fluids in presence of both rotation and uniform external magnetic field have not been studied so far. In this paper, we explore this issue in electrically conducting fluids by simultaneous performance of direct numerical simulations and low dimensional modeling with free-slip perfectly electrically and thermally conducting boundaries. For simplicity, we consider the limits $\mathrm{Pm}, \mathrm{Pr}\rightarrow 0$ and vary the values of $\mathrm{Q}$ and $\mathrm{Ta}$ in the range 
$0 <  \mathrm{Q}, \mathrm{Ta} \leq 50$. Our investigation reveals a rich dynamics near the onset of convection including periodic and chaotic ones at the onset of convection. Origin of chaos and periodic solutions are analyzed in detail using a low dimensional model derived from the DNS data. 


\section{Rotating Hydromagnetic System}
The hydromagnetic system considered here consists of a thin layer of electrically conducting Boussinesq fluid of thickness $d$, thermal diffusivity $\kappa$, magnetic diffusivity $\lambda$, coefficient of volume expansion $\alpha$ and kinematic viscosity $\nu$ kept between two horizontal conducting plates. The system is rotating uniformly with angular velocity $\Omega$ about the vertical axis and heated from below in presence of an external uniform horizontal magnetic field ${\bf B}_0 \equiv (0, B_0, 0)$. The temperatures of the lower and upper plates are $\mathrm{T}_l$ and $\mathrm{T}_u$ respectively $(\mathrm{T}_l > \mathrm{T}_u)$. 
\begin{figure}[h]
\begin{center}
\includegraphics[height=!,width=0.8\textwidth]{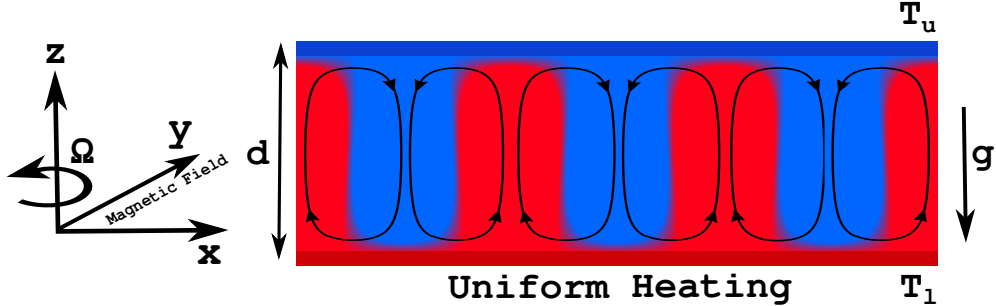}
\end{center}
\caption{Schematic diagram of rotating magnetoconvection. }\label{fig:rbc}
\end{figure}
A schematic diagram of the magnetoconvective set-up has been shown in figure~\ref{fig:rbc}. The nondimensionalized equations for the system under Boussinesq approximation are then given by,
\begin{eqnarray}
\frac{\partial \bf{v}}{\partial t} + (\bf{v}.\nabla)\bf{v} &=& -\nabla{\pi} + \nabla^2{\bf{v}} + \mathrm{Ra} \theta {\bf{\hat{e_3}}}+ \sqrt{\mathrm{Ta}}(\bf{v}\times {\bf{\hat{e_3}}})\nonumber \\ && +\mathrm{Q}\left[\frac{\partial{\bf b}}{\partial y} + \mathrm{Pm}({\bf b}{\cdot}\nabla){\bf b}\right] , \label{eq:momentum} \\   
\mathrm{Pm}[\frac{\partial{\bf b}}{\partial t} + ({\bf v}{\cdot}\nabla){\bf b} &-& ({\bf b}{\cdot}\nabla){\bf v}] = {\nabla}^2 {\bf b} + \frac{\partial{\bf v}}{\partial y},\label{eq:magnetic}\\      
\mathrm{Pr}\left[\frac{\partial \theta}{\partial t}+(\bf{v}.\nabla)\theta\right] &=& v_3+\nabla^2\theta \label{eq:heat},\\
\nabla . \bf{v}=0, && \nabla . \bf{b}=0, \label{eq:div_free}
\end{eqnarray}
where $\mbox{\bf {v}} (x,y,z,t) \equiv (\mathrm{v_1}, \mathrm{v_2}, \mathrm{v_3})$ is the velocity field, $\mbox{\bf{b}} (x,y,z,t) \equiv (\mathrm{b_1},\mathrm{b_2}, \mathrm{b_3})$ is the induced magnetic field, $\theta (x, y, z, t)$ is the deviation in temperature field from the steady conduction profile, $g$ the acceleration due to gravity, and $\beta d$ = $\Delta T$ = $\mathrm{T}_l - \mathrm{T}_u$ is the temperature difference across the fluid layer and the unit vector $\mbox{\boldmath $\hat{e}$}_3$ is directed vertically upward. Non-dimensionalization of the hydrodynamic system is done using the units $\frac{d^2}{\nu}$ for time, $d$ for length, $\frac{\nu}{d}$ for velocity, $\frac{B_0\nu}{\lambda}$ for induced magnetic field, and $\frac{\Delta T \nu}{\kappa}$ for temperature. Non-dimensionalization process gives rise five dimensionless numbers namely the Rayleigh number $\mathrm{Ra}=\frac{\alpha \beta g d^4}{\nu \kappa}$, the Chandrasekhar number $\mathrm{Q}=\frac{{B_0}^2 d^2}{\nu \lambda \rho_0}$, the Taylor number $\mathrm{Ta} = \frac{4\Omega^2d^4}{\nu^2}$, the Prandtl number $\mathrm{Pr}=\frac{\nu}{\kappa}$ and the magnetic Prandtl number $\mathrm{Pm}=\frac{\nu}{\lambda}$. We also use another parameter called reduced Rayleigh number defined by $r = \frac{\mathrm{Ra}}{\mathrm{Ra_c}}$ in the subsequent discussions.

In this paper, we are interested to investigate instabilities and bifurcations in electrically conducting low Prandtl number fluid convection ($\mathrm{Pr}\approx 10^{-2}$) for which the magnetic Prandtl number is also very low ($\mathrm{Pm} \approx 10^{-5}$). The governing equations (\ref{eq:momentum}) - (\ref{eq:div_free}) of the system involve four nonlinear terms and it is difficult to deal with them in low Prandtl number regime due to critical slowing down.  Therefore,  for simplicity here we simultaneously consider $\mathrm{Pr}\rightarrow 0$ and $\mathrm{Pm}\rightarrow 0$. In absence of the magnetic field, $\mathrm{Pr}\rightarrow 0$ limit has already been considered~\cite{thual:1992,kft,pal:2009,priyanka:2014} and the reduced equations revealed significant properties of low Prandtl number convection~\cite{thual:1992,mishra:2010}. In this limit, the equations~(\ref{eq:momentum}),~(\ref{eq:magnetic}) and~(\ref{eq:heat}) reduces to 
\begin{eqnarray}
\frac{\partial \bf{v}}{\partial t} + (\bf{v}.\nabla)\bf{v} &=& -\nabla{\pi} + \nabla^2{\bf{v}} + \mathrm{Ra} \theta {\bf{\hat{e_3}}}\nonumber \\ && + \sqrt{\mathrm{Ta}}(\bf{v}\times {\bf{\hat{e_3}}})+\mathrm{Q}\frac{\partial{\bf b}}{\partial y}, \label{eq:momentum1} \\    
\nabla^2\bf{b} &=& -\frac{\partial \bf{v}}{\partial y} ~\mathrm{and}~ \nabla^2\theta = -v_3.\label{eq:magnetic1} 
\end{eqnarray}
The top and bottom plates are assumed to be stress-free and perfectly thermally conducting which imply 
\begin{equation}
v_3 = \frac{\partial v_1}{\partial z} = \frac{\partial v_2}{\partial z} = \theta = 0~\mathrm{at}~ z = 0, 1.\label{bc1}
\end{equation}
Periodic boundary conditions are assumed in the horizontal directions. The perfectly electrically conducting horizontal boundaries give
\begin{equation}
b_3 = \frac{\partial b_1}{\partial z} = \frac{\partial b_2}{\partial z} = 0~\mathrm{at}~ z = 0, 1.\label{bc2}
\end{equation}
Therefore, the mathematical model of the system under consideration consists of the equations~(\ref{eq:momentum1}) and~(\ref{eq:magnetic1}) together with the boundary conditions~(\ref{bc1}) and~(\ref{bc2}). In the next section we discuss briefly about the linear stability analysis for the onset of convection and compare it with the direct numerical simulation results.
\section{Linear Stability Analysis and comparison with DNS}
We perform linear stability analysis of the conduction state to determine the onset of convection using normal mode analysis following Chandrasekhar~\cite{chandra:book}. In which, we only consider the linearised system and express an arbitrary disturbance as a superposition of certain basic possible modes and examine the stability of the system with respect to each of these modes. The linearised system in our case is given by:
\begin{eqnarray}
\frac{\partial \bf{v}}{\partial t} &=& -\nabla{\pi} + \nabla^2{\bf{v}} + \mathrm{Ra} \theta {\bf{\hat{e_3}}}\nonumber \\ && + \sqrt{\mathrm{Ta}}(\bf{v}\times {\bf{\hat{e_3}}})+\mathrm{Q}\frac{\partial{\bf b}}{\partial y}, \label{eq:momentum_linear} \\    
\nabla^2\bf{b} &=& -\frac{\partial \bf{v}}{\partial y} ~\mathrm{and}~ \nabla^2\theta = -v_3.\label{eq:magnetic_linear} 
\end{eqnarray}
Taking curl of the equation (\ref{eq:momentum_linear}) and considering the vertical component we get, 
\begin{equation}
\frac{\partial \omega_3}{\partial t} = \nabla^2 \omega_3 + \frac{\partial}{\partial y}(\nabla \times {\bf b})\cdot {\bf{\hat{e_3}}} + \sqrt{\mathrm{Ta}} \frac{\partial v_3}{\partial z}, \label{eq:single_curl}
\end{equation}
where, $\omega_3$ is the vertical component of vorticity. Taking curl twice of the equation (\ref{eq:momentum_linear}) we get the following equation
\begin{equation}
\frac{\partial}{\partial t} \nabla^2 v_3 = \nabla^4 v_3 + \mathrm{Ra} \nabla_H^2 \theta - \mathrm{Q} \frac{\partial^2 v_3}{\partial y^2} - \sqrt{\mathrm{Ta}} \frac{\partial \omega_3}{\partial z}, \label{eq:twice_curl}
\end{equation}
for the vertical velocity, where $\nabla_H = \frac{\partial^2}{{\partial x}^2}+\frac{\partial^2}{{\partial y}^2}$ is horizontal laplacian. Eliminating $\theta$ and $\omega_3$ from (\ref{eq:twice_curl}) using (\ref{eq:magnetic_linear}) and (\ref{eq:single_curl}), we get

\begin{eqnarray}
 [\lbrace\nabla^3(\partial_t &-& \nabla^2) \nonumber+ \mathrm{Q}\nabla D_y^2\rbrace^2 
+ \mathrm{Ta} \nabla^4 D^2 \nonumber\\&-& \mathrm{Ra} \nabla_H^2 \lbrace \nabla ^2 (\partial_t - \nabla^2) + \mathrm{Q} D_y^2 \rbrace ]v_3 = 0,\label{eq:combined}
\end{eqnarray}
where $D_y \equiv \partial_y$ and  $D \equiv \partial_z$. 
We consider the expansion of vertical velocity in normal modes as
\begin{eqnarray}
v_3(x,y,z,t) = W(z)exp[i(k_{x}x + k_{y}y) + \sigma t]\label{eq:normal_mode},
\end{eqnarray}
where, $k_x$ and $k_y$ are the wave numbers along $x$ and $y$ direction respectively and $k = \sqrt{k_x^2 + k_y^2}$ is the horizontal wave number. Inserting equation (\ref{eq:normal_mode}) in equation (\ref{eq:combined}), and choosing a trial solution $W(z) = Asin(\pi z)$, which is compatible with our boundary conditions, we arrived at the following stability condition:
\begin{eqnarray}
(\pi^2 &+& k^2)^3[\lbrace(\pi^2 + k^2 + \sigma) + \frac{\mathrm{Q}k_y^2}{(\pi^2 + k^2)}\rbrace^2 +  \frac{\mathrm{Ta} \pi^2}{(\pi^2 + k^2)} ] \nonumber\\ &=& \mathrm{Ra} k^2 [(\pi^2 + k^2)(\pi^2 + k^2 + \sigma) + \mathrm{Q}k_y^2].\label{eq:stability_general}
\end{eqnarray}

\subsection{Stationary convection}
For stationary convection we set $\sigma = 0$ in equation (\ref{eq:stability_general}) to get
\begin{eqnarray}
(\pi^2 &+& k^2)^3[\lbrace(\pi^2 + k^2 ) + \frac{\mathrm{Q}k_y^2}{(\pi^2 + k^2)}\rbrace^2 +  \frac{\mathrm{Ta} \pi^2}{(\pi^2 + k^2)} ] \nonumber\\ &=& \mathrm{Ra} k^2 [(\pi^2 + k^2)(\pi^2 + k^2) + \mathrm{Q}k_y^2].\label{eq:stability_stationary}
\end{eqnarray}

Therefore, the Rayleigh number $\mathrm{Ra(Ta, Q)}$ is given by
\begin{eqnarray}
\mathrm{Ra(Ta, Q)} &=& \frac{(\pi^2 + k^2)\left[(\pi^2 + k^2)^2 + \mathrm{Q}k_y^2\right]}{k^2} \nonumber\\
&+& \frac{\mathrm{Ta} \pi^2 (\pi^2 + k^2)^2}{k^2 \left[(\pi^2 + k^2)^2 + \mathrm{Q} k_y^2\right]}. \label{eq:rac}
\end{eqnarray}
We now numerically find the minimum values of $\mathrm{Ra}$ which is the critical Rayleigh number ($\mathrm{Ra_c}$) for the onset of convection as a function of $\mathrm{Ta}$ and $\mathrm{Q}$ together with the corresponding critical wave number $k_c = \sqrt{k_x^2 + k_y^2}$. In the ranges of $\mathrm{Ta}$ and $\mathrm{Q}$ considered in this paper, we find that $\mathrm{Ra_c}$ is independent of $\mathrm{Q}$ and $k_c$ is a function of $k_x$ only. So at the onset of convection, 2D rolls start to grow linearly.

To verify the linear theory results for the onset of convection, we simulate the equations~(\ref{eq:momentum1})-(\ref{eq:magnetic1}) together with the boundary conditions~(\ref{bc1})-(\ref{bc2}) using a pseudo-spectral code~\cite{mkv:code}. The equation~(\ref{eq:magnetic1}) shows that convective temperature and induced magnetic fields are slaved to velocity fields. In the simulation code, vertical velocity and  vorticity are expanded as
\begin{eqnarray}
v_3 (x,y,z,t)&=& \sum_{l,m,n} W_{lmn}(t)e^{i(lk_cx+mk_cy)}\sin{(n\pi z)},\nonumber\\
\omega_3 (x,y,z,t) &=& \sum_{l,m,n} Z_{lmn}(t)e^{i(lk_cx+mk_cy)}\cos{(n\pi z)},\nonumber\\
\end{eqnarray}
where $l, m, n$ can take non-negative integer values. Horizontal velocity and vorticity components are then derived from the continuity equation. Induced magnetic field and convective temperature field are determined from equation~(\ref{eq:magnetic1}). Random initial conditions have been used for the simulation. The grid resolution for the simulation is taken to be $64 \times 64 \times 64$ and for time advancement fourth order Runge-Kutta scheme is used with time step $\Delta t = 0.001$.
The critical values of the Rayleigh number are now determined from the direct numerical simulations  for some values of the Taylor number in the range $10\leq\mathrm{Ta}\leq 10^2$ using the critical wave numbers obtained from linear theory and are shown in the table~\ref{table:onset}. From the table it is seen that critical Rayleigh numbers computed from linear theory and DNS are very close. 
\begin{center}
\begin{table}[h]
\caption{Onset of convection as obtained from linear theory (LT) and DNS.}\label{table:onset}
\begin{tabular}{|c|c|c|c|c|c|c|}
\hline 
$\mathrm{Ta}$ & 10 & 20 & 30 & 40 & 50 & 100 \\ 
\hline 
$\mathrm{Ra_c}$(LT) & 677.077 & 695.85 & 713.93 & 731.39 & 748.31 & 826.28\\ 
\hline 
$\mathrm{Ra_c}$(DNS) & 677 & 695 & 713 & 730 & 745 & 820\\ 
\hline  
$k_c$ (LT) & 2.270 & 2.315 & 2.357 & 2.397 & 2.434 & 2.594 \\
\hline
\end{tabular} 
\end{table}
\end{center}
Note that the values of $\mathrm{Ra_c}$ and $k_c$ are found to be independent of $\mathrm{Q}$ in the considered range of values of $\mathrm{Q}$ in this paper. 

\subsection{Oscillatory convection}
To determine the conditions for the onset of convection in the form of oscillatory motion (overstability) we put $\sigma = i\omega_1$ in (\ref{eq:stability_general}) and obtain
\begin{eqnarray}
(\pi^2 &+& k^2)^3[\lbrace(\pi^2 + k^2 + i\omega_1) + \frac{\mathrm{Q}k_y^2}{(\pi^2 + k^2)}\rbrace^2 +  \frac{\mathrm{Ta} \pi^2}{(\pi^2 + k^2)} ] \nonumber\\ &=& \mathrm{Ra} k^2 [(\pi^2 + k^2)(\pi^2 + k^2 + i\omega_1) + \mathrm{Q}k_y^2].\label{eq:overstability_general}
\end{eqnarray}
Comparing real and imaginary parts of (\ref{eq:overstability_general}) we get the expressions for $\mathrm{Ra}$ and $\omega_1$ as
\begin{eqnarray}
\mathrm{Ra(Ta, Q)} = \frac{(\pi^2 + k^2)\left[(\pi^2 + k^2)^2 + \mathrm{Q}k_y^2\right]}{k^2} \nonumber\\
+ \frac{\mathrm{Ta} \pi^2 (\pi^2 + k^2)^2\left[(\pi^2 + k^2)^2 + \mathrm{Q}k_y^2\right]}{k^2 \left[\{(\pi^2 + k^2)^2 + \mathrm{Q} k_y^2\}^2 + \omega_1 ^2(\pi^2 + k^2)^2\right]} \label{eq:rao}
\end{eqnarray}
and
\begin{eqnarray}
\omega_1^2 = \frac{\mathrm{Ta}\pi^2(\pi^2 + k^2) - \left[(\pi^2 + k^2)^2 + \mathrm{Q}k_y^2\right]^2}{(\pi^2 + k^2)^2}.\label{eq:omega1_sqr}
\end{eqnarray}
We then numerically find from equations~(\ref{eq:rao}) and~(\ref{eq:omega1_sqr}) that overstability occurs at the onset of convection when $\mathrm{Ta} \geq 548$ for all values of $\mathrm{Q}$. Interestingly, we note that the minimum value of $\mathrm{Ta}$ required for overstable oscillatory convection at the onset does not depend on $\mathrm{Q}$ and it is same as the one obtained by Chandrasekhar in absence of magnetic field ($\mathrm{Q} = 0$)~\cite{chandra:book}.

In the subsequent analysis, we consider $\mathrm{Ta} < 100$ and hence the discussion only corresponds to stationary convection at the onset. Now to investigate the bifurcation structure near the onset of convection and understand the origin of different solutions as a function of parameters, we perform extensive DNS in the parameter regime of our interest and also do low dimensional modeling of the system. The details of low dimensional modeling  has been described in the next section. 

\section{Low dimensional modeling}\label{ldm}
For deriving low dimensional models, we first select large scale modes of vertical velocity and vorticity using theoretical argument as well as DNS data. 
After selecting the modes, the truncated expressions of $v_3$ and $\omega_3$ are determined. Horizontal components of the velocity $v_1$ and $v_2$ are then determined using the equation of continuity for velocity. Projecting the hydrodynamic equations on these modes, we get a set of coupled nonlinear ordinary differential equations which is a low dimensional model for the investigation of the bifurcation structure near the onset of convection. 

Here we note that the modes of the form $W_{lm0}$ in the vertical velocity are not allowed by the boundary condition. Moreover, as argued in~\cite{hirdesh:2013}, the modes of the form $W_{00n}$ in the vertical velocity are also not excited as there is no horizontal mean flow. Now from the linearized equation~(\ref{eq:twice_curl}) it is seen that the linear growth rate of the vertical velocity mode $W_{101}$ will first become positive as the value of $\mathrm{Ra}$ is increased from conduction state and hence this mode must be excited at the onset. Physically this mode represents a 2D rolls pattern along $y$-axis and it is linearly growing. At this stage we assume that the linear growth of the 2D roll mode along $y$-axis is stopped either by transferring energy to the 2D roll mode in the direction of $x$-axis or to the wavy roll mode in the direction of $y$-axis. Note that linear growth rate of the 2D rolls mode along $x$-axis, $W_{011}$ becomes positive next to $W_{101}$ as the value of $\mathrm{Ra}$ is increased further. In the first case the vorticity mode $Z_{110}$ is responsible for the transfer of energy to the 2D rolls along $x$-axis and in the second case $Z_{010}$ make the rolls along $y$-axis wavy.  From DNS with random initial conditions we observe both the cases and accordingly different sets of modes are excited. In this subsection we discuss the mode selection in the first case. So as a first guess we consider the modes $W_{101}$, $W_{011}$ in vertical velocity and $Z_{101}$, $Z_{011}$  and $Z_{110}$ in vertical vorticity. Projecting the hydrodynamic equation on these modes we find a $5$ dimensional model which diverges for supercritical values of the Rayleigh number. Therefore, we consider first nonlinear correction in the truncated expressions of vertical velocity and vorticity.   We find that the modes $W_{112}$, $W_{211}$, $W_{121}$ in vertical velocity and $Z_{112}$, $Z_{211}$, $Z_{121}$ in vertical vorticity are generated in addition to the previously chosen modes.  Now using these $5$ modes in vertical velocity and $6$ modes in vertical vorticity we construct a $11$ mode model which does not diverge but we do not find any qualitative match of the model results with the DNS results. So we need to go for the second nonlinear correction. In this step, additional $16$  modes in vertical velocity namely $W_{103}$, $W_{013}$, $W_{301}$, $W_{031}$, $W_{202}$, $W_{022}$, $W_{123}$, $W_{213}$, $W_{303}$, $W_{033}$, $W_{321}$, $W_{231}$, $W_{233}$, $W_{323}$, $W_{132}$, $W_{312}$ and $20$ modes in vertical vorticity namely $Z_{103}$, $Z_{013}$, $Z_{301}$, $Z_{031}$, $Z_{202}$, $Z_{022}$, $Z_{123}$, $Z_{213}$, $Z_{303}$, $Z_{033}$, $Z_{321}$, $Z_{231}$, $Z_{233}$, $Z_{323}$, $Z_{132}$, $Z_{312}$,  $Z_{310}$, $Z_{130}$, $Z_{020}$, $Z_{200}$ are generated. We then construct a $47$ dimensional model and find that it is giving results which is consistent with the DNS results. However, to arrive at minimum mode model for the bifurcation analysis,  we remove the higher order modes with least contribution to the total energy yet the resulting low dimensional model gives satisfactory results which matches with the DNS results. In this process, we remove $20$ higher order modes and arrive at the minimum mode model for the study. Finally we select the following expression for vertical velocity and vorticity
\begin{eqnarray}
v_3 &=& [W_{101}(t)\cos{k_c x} + W_{011}(t)\cos{k_c y}]\sin{\pi z}\nonumber\\
&+&W_{112}(t)\cos{k_c x}\cos{k_c y}\sin{2\pi z}\nonumber\\
&+&W_{211}(t)\cos{2k_c x}\cos{k_c y}\sin{\pi z}\nonumber\\ 
&+& W_{121}(t)\cos{k_c x}\cos{2k_c y}\sin{\pi z}\nonumber\\
&+&[W_{202}(t)\cos{2k_c x} + W_{022}(t)\cos{2k_c y}]\sin{2\pi z}\nonumber\\
&+& [W_{301}(t)\cos{3k_c x} + W_{031}(t)\cos{3k_c y}]\sin{\pi z}\nonumber\\
&+& [W_{103}(t)\cos{k_c x} + W_{013}(t)\cos{k_c y}]\sin{3\pi z}\\
\mathrm{and}\nonumber\\
\omega_3 &=& [Z_{101}(t)\cos{k_c x} + Z_{011}(t)\cos{k_c y}]\cos{\pi z}\nonumber\\
&+&Z_{110}(t)\sin{k_c x}\sin{k_c y} \nonumber\\
&+& Z_{112}(t)\sin{k_c x}\sin{k_c y}\cos{2\pi z}\nonumber\\
&+& Z_{211}(t)\sin{2k_c x}\sin{k_c y}\cos{\pi z} \nonumber\\
&+& Z_{121}(t)\sin{k_c x}\sin{2k_c y}\cos{\pi z}\nonumber\\
&+&Z_{310}(t)\sin{3k_c x}\sin{k_c y} + Z_{130}(t)\sin{k_c x}\sin{3k_c y} \nonumber\\
&+&[Z_{202}(t)\cos{2k_c x} + Z_{022}(t)\cos{2k_c y}]\cos{2\pi z}\nonumber\\ 
&+& [Z_{301}(t)\cos{3k_c x} + Z_{031}(t)\cos{3k_c y}]\cos{\pi z}\nonumber\\
&+& [Z_{103}(t)\cos{k_c x} + Z_{013}(t)\cos{k_c y}]\cos{3\pi z}\nonumber\\
&+& Z_{020}(t)\cos{2k_c y}+ Z_{200}(t)\cos{2k_c x}.
\end{eqnarray}
Therefore, the low dimensional model we use for the study is $27$ dimensional. It is interesting to point out here that the contribution of these selected modes to the total energy is above $90\%$ in the entire region of the parameter space. Moreover, the selection of modes by energy analysis of the DNS data as has been done in~\cite{nandu:2016} give similar set of modes for the low dimensional model. The low dimensional models are computationally cheap in terms of computer time. Moreover, from the low dimensional model we can compute the unstable solutions along with stable solutions which helps in understanding the origin of different solutions. On the other hand, DNS always gives the stable solutions.   

\section{Results and Discussions}
We perform bifurcation analysis of the $27$ dimensional model using the MATCONT~\cite{dhooge:matcont_2003} software to identify different flow regimes and understand their origin in the considered range of the parameter space.  We also simultaneously carry out DNS to validate the model results. In absence of the external magnetic field ($\mathrm{Q} = 0$), the bifurcation structure in slowly rotating zero Prandtl-number RBC has been explored by Maity and Kumar~\cite{priyanka:2014}. At the onset, they reported periodic bursting (PB) and as the value of the the Rayleigh number increased they sequentially observed chaotic bursting, oscillatory cross rolls (OCR-I \& OCR-II), stationary cross rolls (CR) and stationary squares (SQ) for $\mathrm{Ta} = 10$. On the other hand, for very low Prandtl number fluids ($\mathrm{Pr} = 0.01$), in presence of weak horizontal external magnetic field and in absence of rotation near the onset of convection straight rolls (SR), oscillatory cross rolls (OCR) and stationary cross rolls (CR) patterns have been reported in~\cite{arnab:2014}. In the next subsections, we discuss the bifurcation structures near the onset of convection in simultaneous presence of rotation and magnetic field.  
\begin{figure}
\begin{center}
\includegraphics[height=!, width=0.8\textwidth]{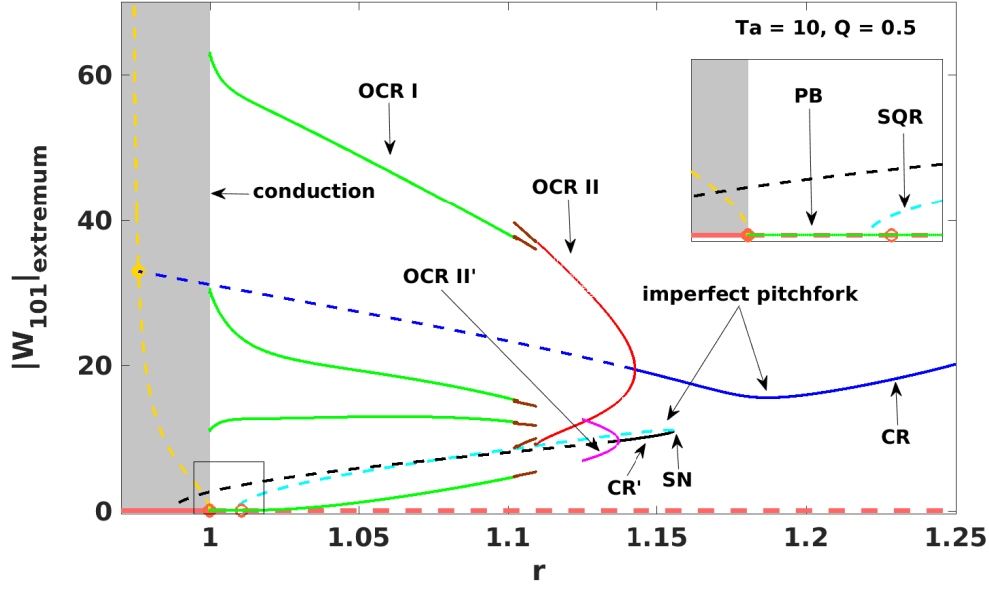}
\end{center}
\caption{Bifurcation diagram computed from the model for $\mathrm{Ta} = 10$ for $\mathrm{Q} = 0.5$.  Extremum values of the variable $|W_{101}|$ for different solutions has been shown as a function of $r$ in the range $0.97\leq r \leq 1.25$ with different colors. Solid and dashed curves represent stable and unstable branches respectively. Orange, yellow, blue and black curves respectively show conduction state, steady 2D rolls along $y$-axis, cross rolls dominant along $y$-axis (CR), cross rolls dominant along $x$ axis (CR$'$). Cyan curve shows the reminiscence of stationary square saddle (SQR). Red and pink curves represent oscillatory cross rolls solutions OCR-II and OCR-II$'$ respectively. Brown and green curves represent different OCR-I type solutions. Orange filled circle shows the subcritical pitchfork bifurcation point at $r = 1$. Filled yellow and empty orange circles respectively show the branch points on 2D rolls and conduction solution branches. Conduction region is shaded with gray color. Saddle-node (SN) and imperfect pitchfork bifurcations are also shown in the diagram. A zoomed view of the marked region is shown at the inset where periodic bursting (PB) and SQR type solutions are shown.}
\label{fig:tau_10_Q_0p5}
\end{figure}
\begin{figure}
\begin{center}
\includegraphics[height=!, width=0.8\textwidth]{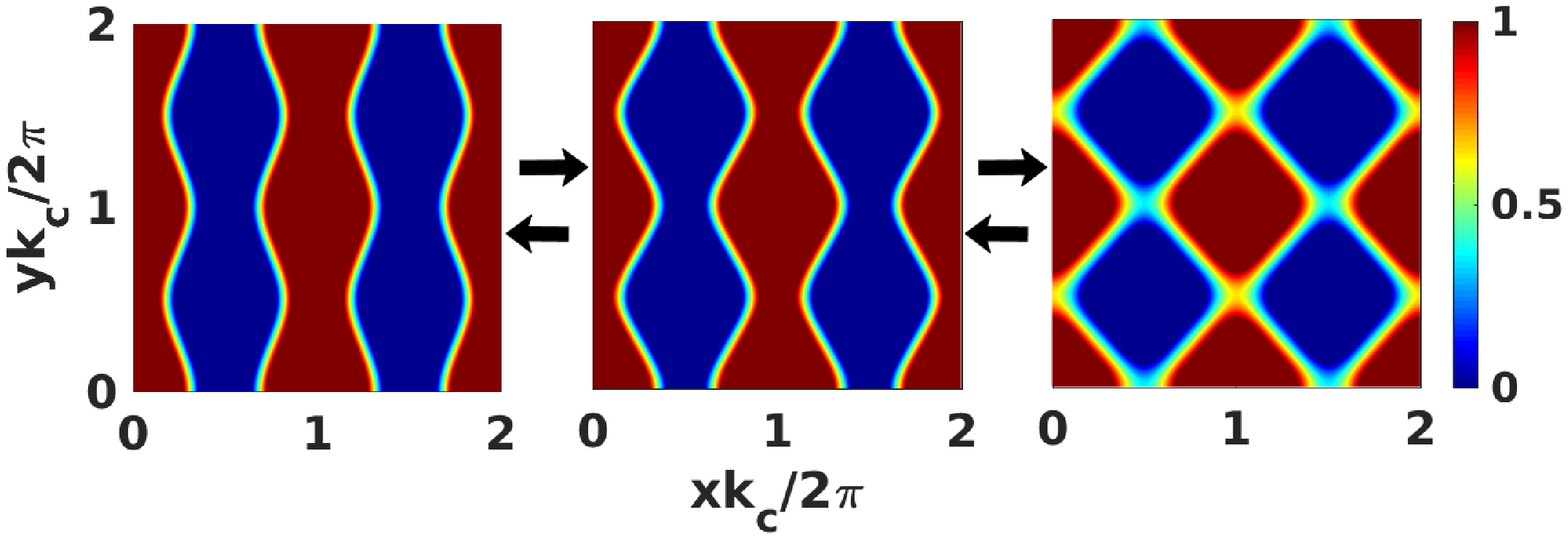}
\end{center}
\caption{Isotherms computed from the model at $z = 0.5$ for $\mathrm{Ta} = 10$, $\mathrm{Q} = 0.5$ and $r = 1.13$ at three different instants showing the pattern dynamics of the OCR-II solution.}
\label{fig:tau_10_Q_0p5_OCRII}
\end{figure}
\begin{figure}[h]
\begin{center}
\includegraphics[height=!, width=0.8\textwidth]{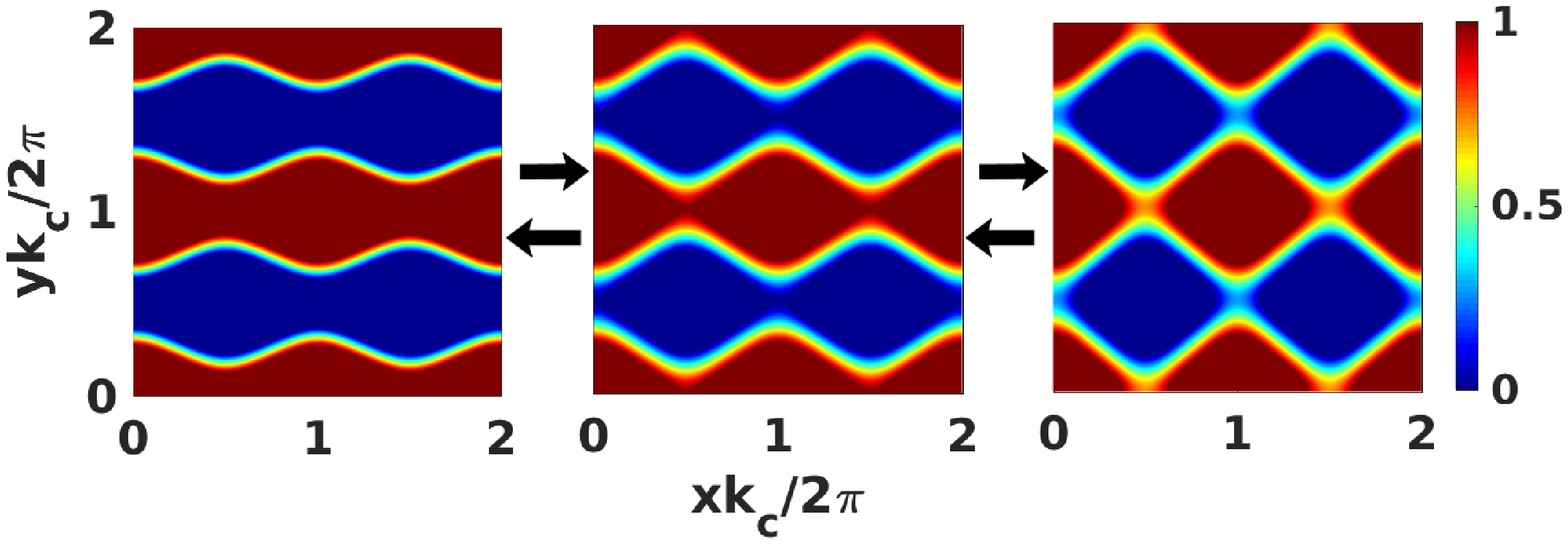}
\end{center}
\caption{Isotherms computed from the model at $z = 0.5$ for $\mathrm{Ta} = 10$, $\mathrm{Q} = 0.5$ and $r = 1.13$ at three different instants showing the pattern dynamics of the OCR-II$'$ solution.}
\label{fig:tau_10_Q_0p5_OCRII'}
\end{figure} 
\begin{figure}
\begin{center}
\includegraphics[height=!, width=0.8\textwidth]{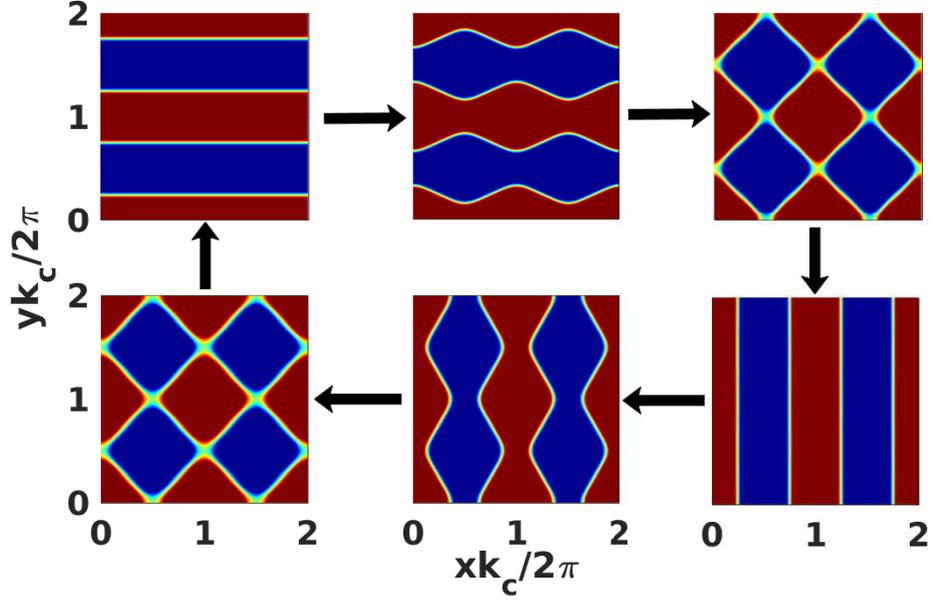}
\end{center}
\caption{Isotherms computed from the model at $z = 0.5$ for $\mathrm{Ta} = 10$ and $\mathrm{Q} = 0.5$ near primary instability $r = 1.004$ at six different instants.}
\label{fig:tau_10_Q_0p5_R_680_pb_pattern}
\end{figure}

\begin{figure*}
\begin{center}
\includegraphics[height=!, width=\textwidth]{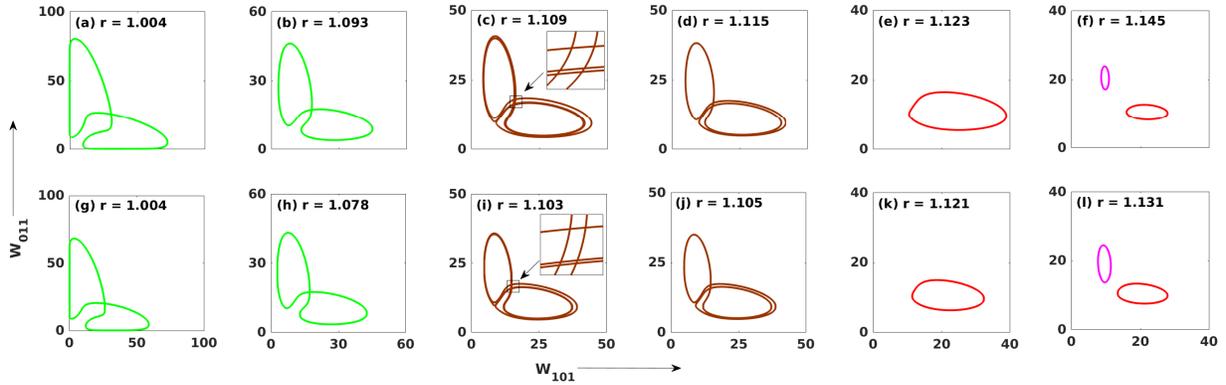}
\end{center}
\caption{Projection of the phase space trajectories on $W_{101} - W_{011}$ plane for $\mathrm{Ta} = 10$, $\mathrm{Q} = 0.5$ for different values of $r$. (a) - (f) are computed from DNS and (g) - (l) are computed using model. Same color coding has been used here as used for similar solutions presented in the figure~\ref{fig:tau_10_Q_0p5}. }
\label{fig:tau_10_Q_0p5_phase_plot}
\end{figure*}

\begin{figure}
\begin{center}
\includegraphics[height=!, width=0.8\textwidth]{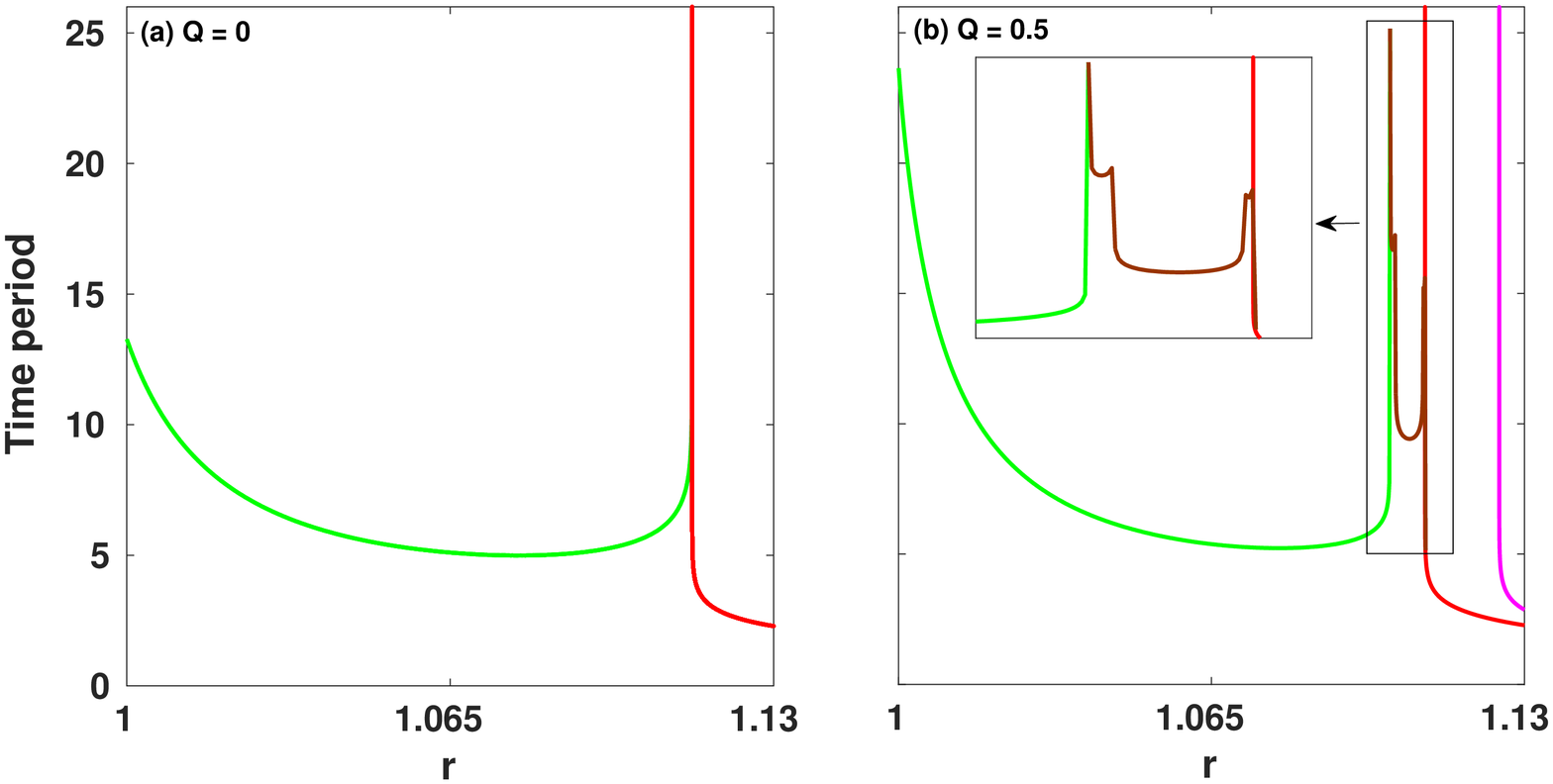}
\end{center}
\caption{Variation of dimensionless time period of OCR-I, OCR-II and OCR-II$'$ solutions with $r$ near the homoclinic bifurcation points for $\mathrm{Ta} = 10$ in absence of magnetic field ($\mathrm{Q} = 0$) and in presence of weak magnetic field ($\mathrm{Q} = 0.5$). (a) Time period of OCR - I and OCR - II solutions are shown with solid green and red curves. (b) Pink and red curves represent time periods of OCR-II and OCR - II$'$ solutions while the brown and green curves show the time periods of OCR-I solutions. An enlarged view of the boxed region is shown at the inset.}
\label{fig:time_period1}
\end{figure}
\subsection{Bifurcation structure for $\mathrm{Q} = 0.5$}
To understand the effect of magnetic field on slowly rotating RBC, we first construct a bifurcation diagram from the model for $\mathrm{Ta} = 10$ and $\mathrm{Q} = 0.5$ (see figure~\ref{fig:tau_10_Q_0p5}). In the bifurcation diagram, extremum values of $W_{101}$ for different solutions together with their stability information have been shown as a function of $r$ in the range $0.97 \leq r \leq 1.25$. The conduction state is stable for $r < 1$ and it becomes unstable via subcritical pitchfork bifurcation at $r = 1$ (filled orange circle) and an unstable 2D rolls branch ($W_{101}\neq 0, W_{011}=0$) is originated. Solid and dashed orange curves in figure~\ref{fig:tau_10_Q_0p5} represent stable and unstable conduction states respectively. The unstable 2D rolls branch is shown with dashed yellow curve. This unstable 2D rolls branch exists in the conduction regime only and an unstable cross rolls (CR) branch (dashed blue curve) is originated from a branch point (filled yellow circle) of it. The unstable CR branch becomes stable (solid blue curve) through an inverse Hopf bifurcation  at $r = 1.142$ and continue to exist till $r = 1.25$ as the value of $r$ is increased. Note that for CR solutions $W_{101}\neq 0, W_{011}\neq 0 ~\mathrm{and}~ W_{101} > W_{011}$. Subcriticality at the onset of convection is an interesting result, the possibility of which was reported long ago by Veronis~\cite{veronis:1959} and recently numerically confirmed in~\cite{knobloch:2013} in rotating convection with low rotation rate. Subcritical thermal convection has also been studied very recently in~\cite{kaplan:2017} in rapidly rotating system under spherical geometry.
\begin{figure}
\begin{center}
\includegraphics[height=!, width=0.8\textwidth]{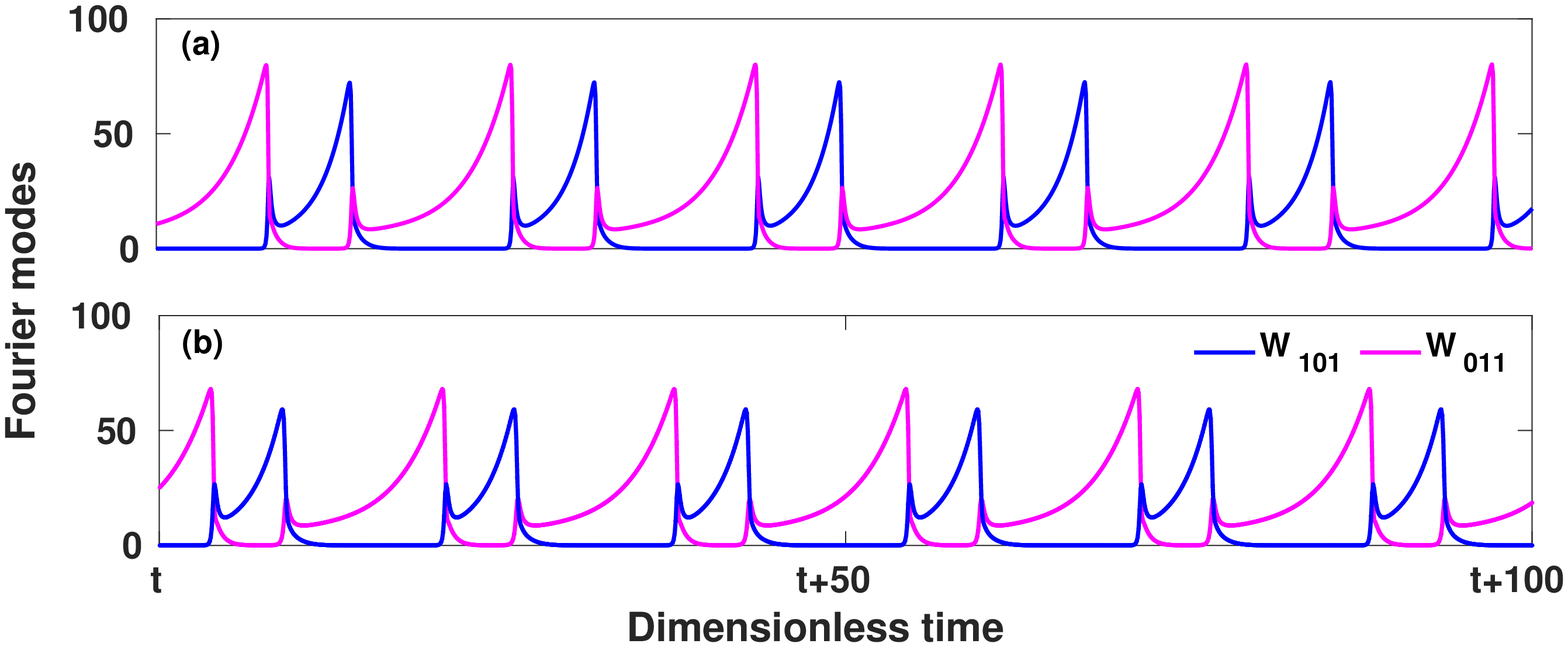}
\end{center}
\caption{Temporal variation of two largest Fourier modes $W_{101}$ (blue curves) and $W_{011}$ (pink curves) near primary instability ($r = 1.004$) for $\mathrm{Ta} = 10$ and $\mathrm{Q} = 0.5$: (a) from DNS and (b) from Model. }
\label{fig:tau_10_Q_0p5_R_680_pb}
\end{figure}
\begin{center}
\begin{table*}
\caption{Flow regimes obtained from DNS and model for $\mathrm{Q} = 0.5$ and different values of the Taylor number $\mathrm{Ta}$  as a function of $r$.}\label{table:flow_pattern_Q_0p5}
\setlength{\tabcolsep}{0.04cm} 
{\renewcommand{\arraystretch}{1}
\begin{tabular}{@{\extracolsep{1pt}}l|cc|cc|cc@{}}
\firsthline 
\hline
Fluid patterns    & DNS & Model & DNS & Model & DNS & Model \\
& $r$($\mathrm{Ta} = 10$) & $r$($\mathrm{Ta} = 10$) & $r$($\mathrm{Ta} = 20$) & $r$($\mathrm{Ta} = 24$) & $r$($\mathrm{Ta} = 30$) & $r$($\mathrm{Ta} = 38$)\\
\hline
Homoclinic Chaos      &  ---    &  ---  & 1.001 - 1.068 & 1.001 - 1.06 & ---  &  ---      \\
OCR-I      & 1.001 - 1.118    & 1.001 - 1.109  & 1.069 - 1.134 & 1.061 - 1.118 & ---  & ---      \\
OCR-II      & 1.119 - 1.149    & 1.110 - 1.142  & 1.135 - 1.160 & 1.119 - 1.148 & 1.001 - 1.165 & 1.001 - 1.150      \\
OCR-II$'$      & 1.133 - 1.145    & 1.124 - 1.136  & 1.148 - 1.156 & 1.138 - 1.144 & ---  & ---      \\
CR$'$      & 1.146 - 1.157    & 1.137 - 1.156   & 1.157 - 1.168 & 1.145 - 1.152 & ---  & ---      \\
CR      &  $\leq$ 1.340  &  $\leq$ 1.275  & $\leq$ 1.325 & $\leq$ 1.253 & $\leq$ 1.314  &  $\leq$ 1.233      \\
\lasthline
\end{tabular}}
\end{table*}
\end{center}
Another unstable 2D rolls branch ($W_{101} = 0, W_{011} \neq 0$) is originated from a branch point of the unstable conduction solution branch at $r = 1.0107$ (open orange circle). A clearer view of this branch point is shown at the inset of the figure~\ref{fig:tau_10_Q_0p5}. The unstable cross rolls (CR$'$) branch is originated at a branch point of this 2D rolls branch (dashed cyan curve) at $r = 1.009$. Note that this branch of unstable 2D rolls solutions is not shown in the bifurcation diagram as for this solution $W_{101} = 0$. Now the CR$'$ branch for which $W_{101}\neq 0, W_{011}\neq 0 ~\mathrm{and}~ W_{101} < W_{011}$ shows a saddle node (SN) bifurcation at $r = 1.1557$ as the value of $r$ is increased and meets with a stable CR$'$ branch (solid black curve) there. As the value of $r$ is now decreased, the stable CR$'$ branch undergoes a supercritical Hopf bifurcation at $r = 1.137$, a stable limit cycle is generated. The CR$'$ branch becomes unstable and continue to exist till very low values of the reduced Rayleigh number (dashed black curve). 

Now if we look at the bifurcation diagram from right hand side, a scenario of imperfect pitchfork bifurcation is clear near $r = 1.187$.  The imperfection is due to the presence external uniform magnetic field in the horizontal direction which breaks the $x \leftrightharpoons y$ symmetry of the problem. In presence of the magnetic field, CR and CR$'$ branches are not connected by the symmetry of the problem. In absence of magnetic field ($\mathrm{Q} =0$), stable CR and CR$'$ branches would meet at a supercritical pitchfork bifurcation point~\cite{pal:2009,pal:2013,priyanka:2014}. Comparing the magnetic and non-magnetic cases we understand that the dashed cyan curves are the reminiscence of the square saddle which would exist when $\mathrm{Q} = 0$ and we denote it by SQR.

As mentioned earlier that CR and CR$'$ branches undergo supercritical Hopf bifurcations at $r = 1.142$ and $r = 1.137$ respectively and stable limit cycles are generated. The limit cycles generated via Hopf bifurcation of the CR and CR$'$ branches show oscillatory cross rolls patterns oriented along $y$-axis (see figure~\ref{fig:tau_10_Q_0p5_OCRII}) and $x$-axis (see figure~\ref{fig:tau_10_Q_0p5_OCRII'}) respectively. We denote these solutions respectively by OCR-II and OCR-II$'$ (solid red and pink curves in the figure~\ref{fig:tau_10_Q_0p5}). Note that OCR-II and OCR-II$'$ solutions are not connected by the symmetry of the problem but exist independently. As the value of $r$ is reduced further, OCR-II$'$ limit cycle becomes homoclinic to SQR saddle at $r = 1.1248$ and ceased to exist. While the OCR-II limit cycle continues to exist and it becomes homoclinic to the SQR saddle at $r = 1.11$. 

Immediately after this homoclinic bifurcation, subsequent reduction of the value of the reduced Rayleigh number shows asymmetric glued limit cycles which continue to exist till $r = 1$. This is an example of imperfect gluing bifurcation which has been reported earlier in electronic circuit~\cite{glendinning:2001}, dynamically complicated extended flow~\cite{abshagen:2001} and Taylor-Couette flow~\cite{marques:2002}. Physically glued limit cycles represent periodic oscillation of rolls (cross rolls) patterns oriented along $x$-axis and $y$-axis with intermediate square patterns. The pattern dynamics for this solution has been shown in the figure~\ref{fig:tau_10_Q_0p5_R_680_pb_pattern}.  These solutions are called oscillatory cross rolls solutions and denoted by OCR-I. The projections of the phase space trajectories on $W_{101}-W_{011}$ plane for different oscillatory cross rolls solutions obtained from DNS and model has been shown in figure~\ref{fig:tau_10_Q_0p5_phase_plot}.
Figures~\ref{fig:tau_10_Q_0p5_phase_plot}(f) and (l) show the phase portraits for coexisting OCR-II (red) and OCR-II$'$ (pink) solutions computed from DNS and model respectively. Flow patterns for these solutions have already been shown in figures~\ref{fig:tau_10_Q_0p5_OCRII} and ~\ref{fig:tau_10_Q_0p5_OCRII'}. For a lower values of $r$, there is a range of $r$ in which only stable OCR-II solutions exist (see figures~\ref{fig:tau_10_Q_0p5_phase_plot}(e) and (k)). As the OCR-II solutions become homoclinic to the SQR saddle at $r = 1.11$, an imperfect gluing bifurcation is observed. Bigger limit cycles of shape $`8`$ encircling the previously existing separate limit cycles are formed. Very close to the bifurcation point, complicated glued limit cycles are observed which encircles one or both the previously existing limit cycles more than once (see figures~\ref{fig:tau_10_Q_0p5_phase_plot}(d), (c), (i) and (j)). Simpler glued limit cycles are observed for lower values of $r$ (figures~\ref{fig:tau_10_Q_0p5_phase_plot}(a), (b), (g) and (h)).     
Pattern dynamics corresponding to the glued solutions are similar to the ones shown in figure~\ref{fig:tau_10_Q_0p5_R_680_pb_pattern}. Note that different types of glued limit cycle solutions are the results of a series of homoclinic gluing bifurcations occurring in the range $1.102\leq r \leq 1.11$ which is evident from divergence of the time period of these solutions (see figure~\ref{fig:time_period1}). From the figures~\ref{fig:tau_10_Q_0p5_phase_plot}(b) and (h) we observe  that as the simple glued limit cycles are originated, the minimum values of $W_{101}$ and $W_{011}$ are away from zero. However, as $r\rightarrow 1$, the minimum values of $W_{101}$ and $W_{011}$ tend to zero and from the time series of these modes shown in figure~\ref{fig:tau_10_Q_0p5_R_680_pb} it is clear that the modes spend significant time near their minimum values before bursting to a large amplitude. This is typical to periodic bursting of flow patterns which is also observed in rotating convection ($\mathrm{Q} = 0$)~\cite{priyanka:2014}.
  
The details of different flow regimes obtained from DNS and model for $\mathrm{Ta} = 10$ and $\mathrm{Q} = 0.5$ have been shown in the second and third columns of the table~\ref{table:flow_pattern_Q_0p5}. The table also shows different flow regimes as the value of $\mathrm{Ta}$ increased for $\mathrm{Q} = 0.5$. Imperfect gluing also observed for $\mathrm{Ta} = 20$ in DNS and near the onset of convection homoclinic chaos is observed. With further increase in the value of $\mathrm{Ta}$, gluing bifurcation is not observed any more. The CR$'$ solution and the class of solutions originated from it are suppressed.  Only the class of solutions originated from CR branch are observed near the onset of convection and very close to the onset of convection, oscillatory cross rolls solutions are observed. However, for slightly higher values of $\mathrm{Ta} ~(\sim 50)$, we observe intermittent chaos at the onset of convection and the intermittency is found to be of type-I. 
\begin{figure}
\begin{center}
\includegraphics[height=!, width=0.8\textwidth]{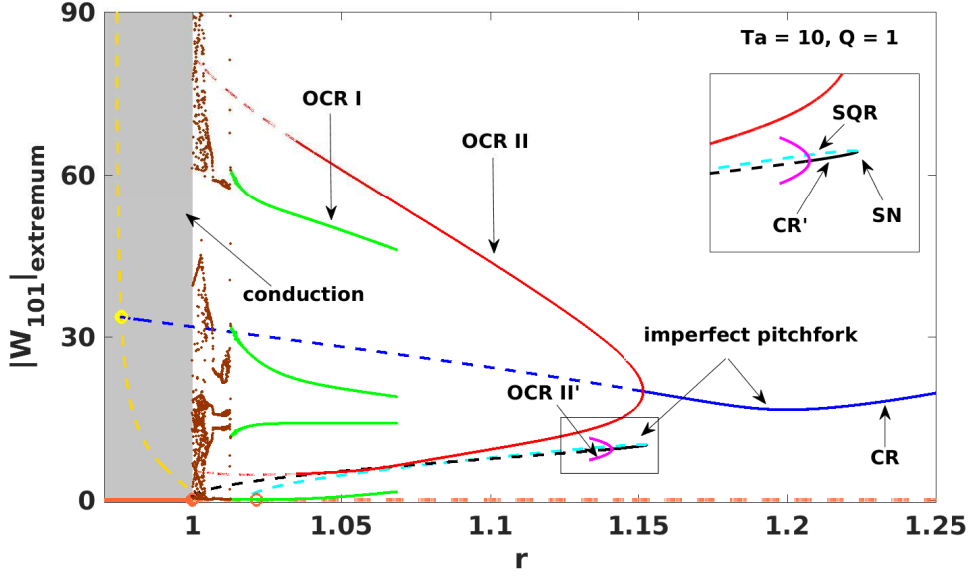}
\end{center}
\caption{Bifurcation diagram constructed from the model for $\mathrm{Ta} = 10$ for $\mathrm{Q} = 1$ in the range $0.97\leq r \leq 1.25$. Solid and dashed curves represent stable and unstable branches respectively. Orange, yellow, blue and black curves respectively show conduction state, steady 2D rolls along $y$-axis, cross rolls along $y$-axis (CR), cross rolls along $x$ axis (CR$'$). Cyan curve shows the reminiscence of stationary square saddle (SQR). Red and pink curves represent oscillatory cross rolls solutions OCR-II and OCR-II$'$ respectively. Green curves represent OCR-I type solutions. Orange filled circle shows the subcritical pitchfork bifurcation point at $r = 1$. Filled yellow and empty orange circles respectively show the branch points on 2D rolls and conduction solution branches. Conduction region is shown with shaded gray color. The brown dots represent chaotic solutions near the onset of convection. An enlarged view of the marked region is shown at the inset where stationary cross rolls (CR$'$), SQR type solutions and saddle-node (SN) bifurcation point have been displayed.}
\label{fig:tau_10_Q_1}
\end{figure}
\subsection{Bifurcation structure for $\mathrm{Q} = 1$}
Now we construct a bifurcation diagram from the model for $\mathrm{Ta} = 10$ and $\mathrm{Q} = 1$ (see figure~\ref{fig:tau_10_Q_1}). From the bifurcation diagram we observe that slight change in the strength of the magnetic field brings about a substantial change in the flow structure near the onset of convection. The origin of the CR and CR$'$ solutions are same as it has been for $\mathrm{Q} = 0.5$ i.e. originated from the branch points of two different unstable 2D rolls branches. OCR-II and OCR-II$'$ are generated from the supercritical Hopf bifurcations of the CR and CR$'$ branches respectively. OCR-II$'$ limit cycle then becomes homoclinic to the SQR saddle at $r = 1.133$ but the OCR-II branch never becomes homoclinic to it although it comes very close to the SQR saddle. However, asymmetric glued solutions (OCR-I) are found to co-exist with the OCR-II solutions in a range of $r$ around $1.05$. OCR-I is then becomes homoclinic to the SQR saddle and homoclinic chaos is observed at the onset. 
\begin{figure}
\begin{center}
\includegraphics[height=!, width=0.8\textwidth]{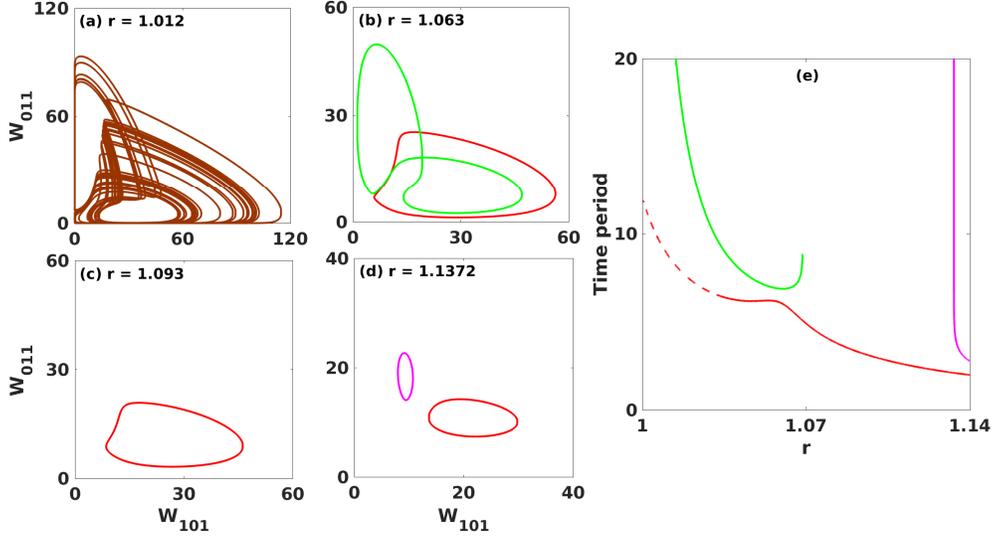}
\end{center}
\caption{(a)-(d) Projections of phase space trajectories on $W_{101} - W_{011}$ plane, for $\mathrm{Ta} = 10$, $\mathrm{Q} = 1$, for different values of $r$. (e) The time periods of OCR-I (solid green curve), OCR-II$'$ (solid pink curve) and OCR-II (red curve) solutions.}
\label{fig:tau_10_Q_1_phase_plot}
\end{figure}

\begin{figure}
\includegraphics[height = !, width=0.8\textwidth]{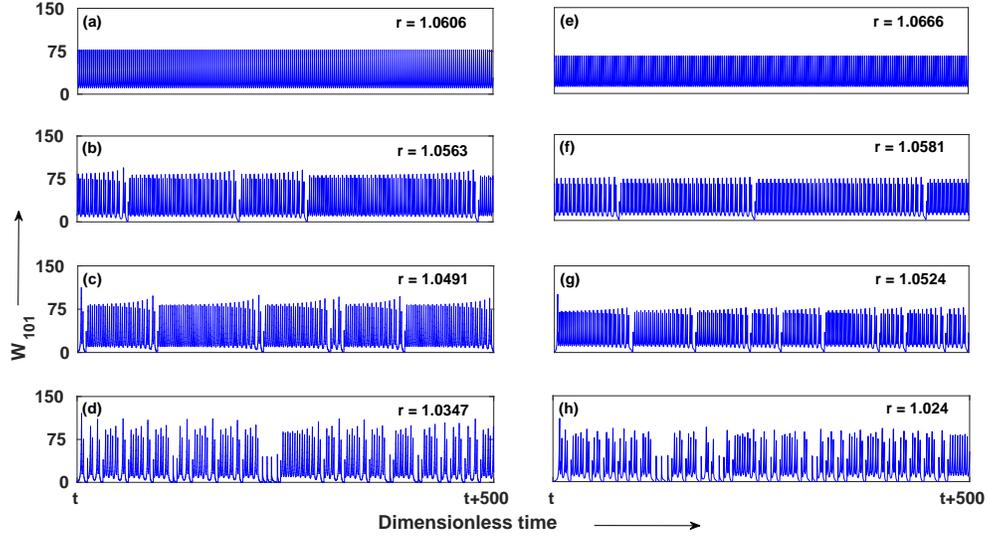}
\caption{Time series of the Fourier mode $W_{101}$ computed from DNS and model for four different values of $r$ to show the intermittent route to chaos. (a) - (d) Computed from DNS for $\mathrm{Ta} = 20$ and $\mathrm{Q} = 1$. (e) - (h) Computed from the model for $\mathrm{Ta} = 24$ and $\mathrm{Q} = 1$.}
\label{fig:intermittency_time_series}
\end{figure}
\begin{figure}
\includegraphics[height = !, width=0.8\textwidth]{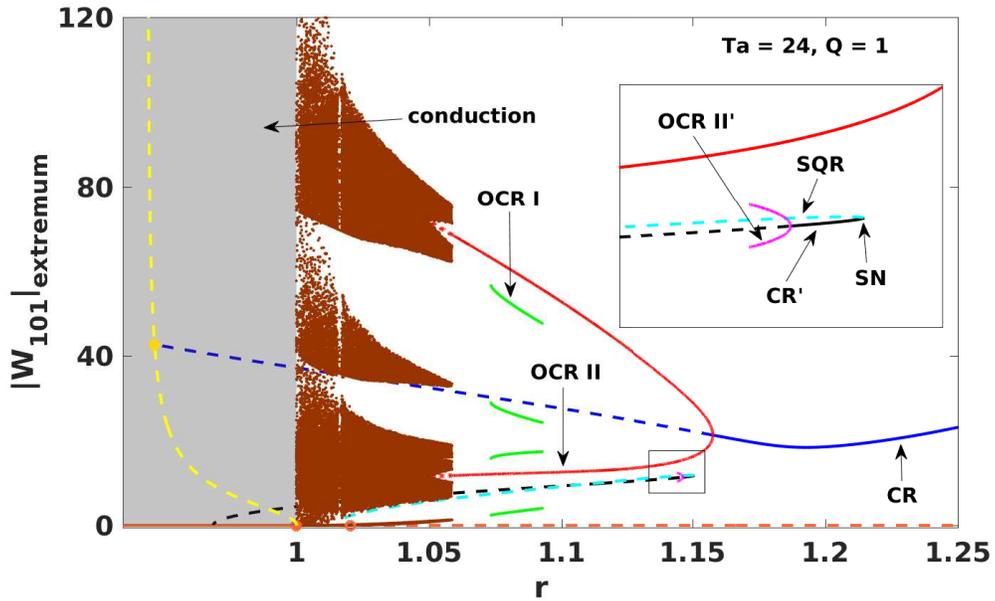}
\caption{Bifurcation diagram constructed form the model for $\mathrm{Ta} = 24$ and $\mathrm{Q} = 1$. Same color coding has been used here for different solutions as it has been used in the previous bifurcation diagrams. The brown dots show the chaotic solutions appeared via type-III intermittency.}
\label{fig:tau_24_Q_1}
\end{figure}

\begin{center}
\begin{table*}
\caption{Flow regimes obtained from DNS and model for $\mathrm{Q} = 1$ and different values of the Taylor number $\mathrm{Ta}$  as a function of $r$.}\label{table:flow_pattern_Q_1}
\setlength{\tabcolsep}{0.01cm} 
{\renewcommand{\arraystretch}{1}
\begin{tabular}{@{\extracolsep{1pt}}l|cc|cc|cc@{}}
\firsthline 
\hline
Fluid patterns    & DNS & Model & DNS & Model & DNS & Model \\
& $r$($\mathrm{Ta} = 10$) & $r$($\mathrm{Ta} = 10$) & $r$($\mathrm{Ta} = 20$) & $r$($\mathrm{Ta} = 24$) & $r$($\mathrm{Ta} = 30$) & $r$($\mathrm{Ta} = 38$)\\
\hline
Homoclinic Chaos      &  1.001 - 1.016    &  1.001 - 1.012  & --- & --- & ---  &  ---      \\
Intermittent Chaos      &  ---    &  ---  & 1.001 - 1.057 & 1.001 - 1.058 & ---  & ---      \\
OCR-II (Period two)      &  ---    &  ---  & --- & --- & 1.001 - 1.058  & 1.001 - 1.096     \\
OCR-I      & 1.017 - 1.071    & 1.013 - 1.068  & 1.086 - 1.102 & 1.073 - 1.092 & ---  & ---      \\
OCR-II      & 1.072 - 1.158    & 1.069 - 1.152  & 1.103 - 1.168 & 1.093 - 1.156 & 1.059 - 1.175  & 1.097 - 1.156      \\
OCR-II$'$      & 1.142 - 1.151    & 1.133 - 1.141  & 1.157 - 1.160 & 1.145 - 1.146 & ---  & ---      \\
OCR-II$''$      & ---    & ---  & 1.058 - 1.085 & 1.059 - 1.072 & ---  & ---      \\
CR$'$      & 1.152 - 1.154    & 1.142 - 1.153   & 1.161 - 1.164 & 1.146 - 1.151 & ---  & ---      \\
CR      &  $\leq$ 1.347  &  $\leq$ 1.282  & $\leq$ 1.334 & $\leq$ 1.262 & $\leq$ 1.324  &  $\leq$ 1.24      \\
\lasthline
\end{tabular}}
\end{table*}
\end{center}
The projections of the phase space trajectories of for the OCR-II, OCR-II$'$, OCR-I and chaotic solutions have been shown in the figures~\ref{fig:tau_10_Q_1_phase_plot}(a), (b), (c) and (d). Figure~\ref{fig:tau_10_Q_1_phase_plot}(e) shows the variation of time period of the OCR-II$'$, OCR-II and OCR-I solutions with pink, red and green colors. It is observed that time period of the OCR-II$'$ solutions diverge near the homoclinic bifurcation point at $r = 1.133$. The time period of the OCR-I solution also diverges as it becomes homoclinic to SQR at $r = 1.014$ and homoclinic chaos is observed near the onset of convection.  Interestingly, chaos at the onset of convection via intermittency route is observed for little higher values of the Taylor number. The time series of the Fourier amplitude $W_{101}$ corresponding to  intermittent route to chaos have been shown in figure~\ref{fig:intermittency_time_series} as obtained from DNS and model for $\mathrm{Ta} = 20$ and $24$ respectively. Note that intermittency here is of type-III. The bifurcation diagram shown in figure~\ref{fig:tau_24_Q_1} shows the chaotic solutions appearing though intermittency with brown dots.  If the value of $\mathrm{Ta}$ is increased a bit ($\mathrm{Ta}\sim 30$), part of the bifurcation structure associated with CR$'$ is suppressed and only OCR-II type solutions are observed at the onset of convection. This scenario is maintained till $\mathrm{Ta} = 50$. Table~\ref{table:flow_pattern_Q_1} shows the detailed flow regimes as obtained from DNS and model for three values of $\mathrm{Ta}$ with $\mathrm{Q} = 1$. Similar bifurcation structures are observed near the onset of convection as the value of $\mathrm{Ta}$ is varied in the range $0 <  \mathrm{Ta} \leq 50$ and $\mathrm{Q} < 2$. Significant change in the flow structure occurs for $\mathrm{Q} \geq 2$ as the value of $\mathrm{Ta}$ is varied in the same range which is discussed in the nest section.

\subsection{Bifurcation structure for $\mathrm{Q} \geq 2$}
\begin{figure}
\includegraphics[height = !, width=0.8\textwidth]{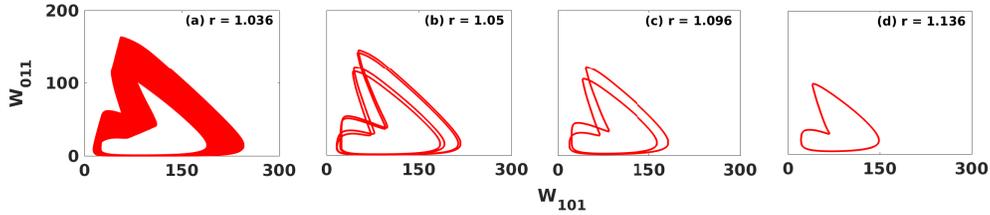}
\caption{Projection of phase space trajectories on $W_{101} - W_{011}$ plane for $\mathrm{Ta} = 50$, $\mathrm{Q} = 2$, for different values of $r$, are computed using model showing peroid doubling route to chaos at the onset of convection.}
\label{fig:tau_50_Q_2}
\end{figure}

\begin{figure}
\includegraphics[height = !, width=0.8\textwidth]{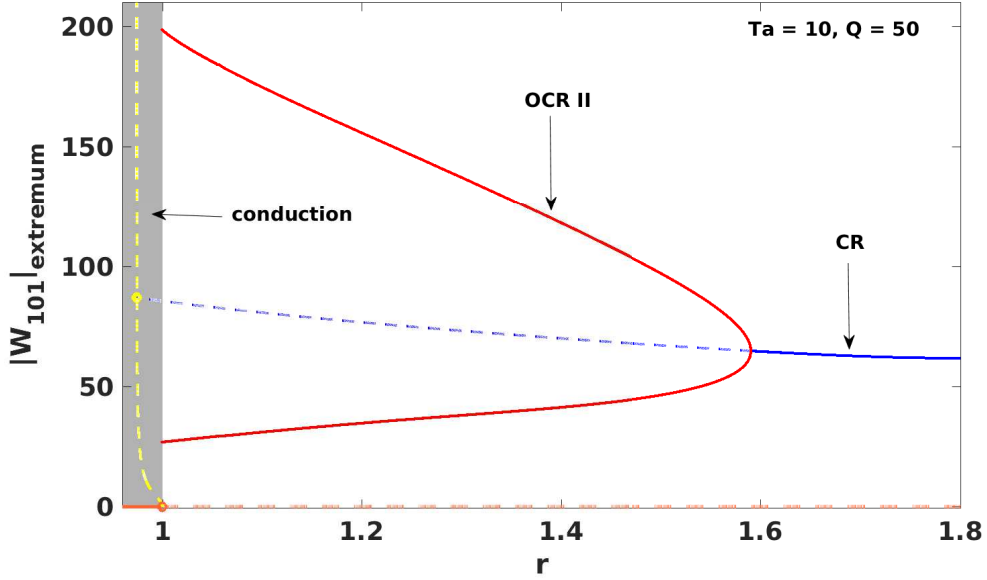}
\caption{Bifurcation diagram constructed from the model for $\mathrm{Ta} = 10$ and $\mathrm{Q} = 50$ in the range $0.97\leq r \leq 1.25$. Solid and dashed curves represent stable and unstable branches respectively. Orange, yellow, and blue curves respectively show conduction state, steady 2D rolls along $y$-axis, and cross rolls along $y$-axis (CR). Red curves represent oscillatory cross rolls solutions (OCR-II). Filled yellow circle show the branch point on 2D rolls solution branch. Conduction region is shown with shaded gray color.}
\label{fig:tau_10_Q_50}
\end{figure}

For $\mathrm{Q}\geq 2$, as the value of $\mathrm{Ta}$ is varied in the range $0 < \mathrm{Ta}\leq 50$ at the onset either oscillatory cross rolls or chaotic solutions are observed but gluing bifurcation is not observed. For lower values of $\mathrm{Ta}$, only periodic oscillatory convection is found to occur at the onset. On the other hand chaotic solutions are found to appear at the onset for higher values of $\mathrm{Ta}$ and interestingly the routes to chaos at the onset is period doubling in this case. Figure~\ref{fig:tau_50_Q_2} shows sequence of projections of the phase space trajectories on the $W_{101}-W_{011}$ plane corresponding to such a period doubling route to chaos for $\mathrm{Ta} = 50$ and $\mathrm{Q} = 2$ obtained from model. Detailed flow regimes obtained from DNS and model for $\mathrm{Q}=2$ are shown in the table~\ref{table:flow_pattern_Q_2}. For higher values of $\mathrm{Q}$, if we set the value of Taylor number in the range $0< \mathrm{Ta}\leq 50$, only oscillatory cross rolls solutions of the type OCR-II are observed at the onset. Figure~\ref{fig:tau_10_Q_50} shows a bifurcation diagram computed from the model for $\mathrm{Q}=50$ and $\mathrm{Ta} = 10$. In the bifurcation diagram, different solutions are shown with different colors with their stability information in the reduced Rayleigh number range $0.97\leq r\leq 1.8$. 
\begin{center}
\begin{table*}
\caption{Flow regimes obtained from DNS and model for $\mathrm{Q} = 2$ and different values of the Taylor number $\mathrm{Ta}$  as a function of $r$.}\label{table:flow_pattern_Q_2}
\setlength{\tabcolsep}{0.01cm} 
{\renewcommand{\arraystretch}{1}
\begin{tabular}{@{\extracolsep{1pt}}l|cc|cc|cc@{}}
\firsthline 
\hline
Fluid patterns    & DNS & Model & DNS & Model & DNS & Model \\
& $r$($\mathrm{Ta} = 10$) & $r$($\mathrm{Ta} = 10$) & $r$($\mathrm{Ta} = 20$) & $r$($\mathrm{Ta} = 24$) & $r$($\mathrm{Ta} = 30$) & $r$($\mathrm{Ta} = 38$)\\
\hline
Chaos      & ---    &  ---  & --- & 1.001 - 1.021 & ---  &  ---      \\
OCR-II (Period two)      &  1.001 - 1.013  &  1.001 - 1.014  & 1.001 - 1.033 & 1.022 - 1.055 & ---  & 1.001 - 1.111   \\
OCR-II      & 1.014 - 1.174   & 1.015 - 1.170  & 1.034 - 1.184 & 1.056 - 1.175 & 1.001 - 1.191  & 1.112 - 1.174      \\
CR      &  $\leq$ 1.362  &  $\leq$ 1.30  & $\leq$ 1.365 & $\leq$ 1.277 & $\leq$ 1.34  &  $\leq$ 1.255     \\
\lasthline
\end{tabular}}
\end{table*}
\end{center}
The solid orange curve represent the stable conduction state for $r < 1$ which becomes unstable via a subcritical pitchfork bifurcation at $r = 1$ and unstable 2D rolls branches are originated there. One of the unstable 2D rolls branch has been shown with dashed yellow curve. Unstable conduction states continue to exist for higher values of $r$ and have been shown with dashed orange curve in the bifurcation diagram. An unstable CR branch is generated from a branch point (filled yellow circle in the bifurcation diagram) of the unstable 2D rolls solution. The CR branch later becomes stable through an inverse Hopf bifurcation at $r = 1.5907$. The Hopf bifurcation is supercritical and stable limit cycle is generated there. These limit cycles physically represent OCR-II type solutions and have been shown with solid red curves in the bifurcation diagram. The CR$'$ solution branch and the associated bifurcation structure are suppressed for higher values of $\mathrm{Q}$. Note that the model under consideration is valid for $\mathrm{Q}\leq 70$.

\subsection{Convective heat flux}
Changes in global quantity like Nusselt-number ($\mathrm{Nu}$) is often considered as the characterization of convective system, which is the ratio of total heat flux (conductive and convective) to conductive heat flux. For the low Prandtl-number fluid convection the expression for Nusselt number is given by~\cite{thual:1992}
\begin{equation}
\mathrm{Nu} = 1 + \mathrm{Pr}^2\langle v_3\theta \rangle,
\end{equation} 
where $\langle\cdot\rangle$ denotes the volume average. Now, in the limit $\mathrm{Pr}\rightarrow 0$, $\mathrm{Nu}$ becomes $1$. Physically this implies, mean profile of temperature in the layer is equivalent to conductive profile~\cite{thual:1992}. However, $\langle v_3 \theta\rangle$ is an interesting quantity (gives convective heat flux~\cite{mkv:2017}) to measure in the limit $\mathrm{Pr}\rightarrow 0$. So here we have computed the quantity $\langle v_3\theta\rangle$ from direct numerical simulation data. Table ~\ref{table:convective_heat_flux} shows the variation of convective heat flux in different flow regimes. First two rows of the table correspond to $\mathrm{Ta} =10$ which show that for a fixed value of $\mathrm{Ta}$ as the value of $\mathrm{Q}$ increases average convective heat flux also increases. Remaining rows of the table correspond to $\mathrm{Q} =1$ which reveal that as the value of $\mathrm{Ta}$ is increased for a fixed value of $\mathrm{Q}$, convective heat flux increases. 
\begin{center}
\begin{table*}
\caption{Measure of convective heat flux in different flow regimes computed from DNS for different values of $\mathrm{Q}$ with $\mathrm{Ta} = 10$ (first two rows) and for different values of  $\mathrm{Ta}$ with $\mathrm{Q} = 1$ (last two rows). }\label{table:convective_heat_flux}
\setlength{\tabcolsep}{0.01cm} 
{\renewcommand{\arraystretch}{1}
\begin{tabular}{@{\extracolsep{1pt}}ccc|ccc@{}}
\firsthline 
\hline
   & $\mathrm{Ta} = 10$ and $\mathrm{Q} = 0.5$ &  &  & $\mathrm{Ta} = 10$ and $\mathrm{Q} = 1$ &  \\
 $r$ & $\langle v_3 \theta\rangle$ & Fluid patterns & $r$ & $\langle v_3 \theta\rangle$ & Fluid patterns\\
\hline
 1.05   &  2.19 - 56.86  &  OCR-I   &  1.05   &  1.4 - 65.9    &  OCR-I     \\
 1.10   &  2.71 - 33.46  &  OCR-I   &  1.10   &  2.35 - 46.39  &  OCR-II   \\
 1.15   &  7.79 - 11.9   &  OCR-II  &  1.15   &  5.95 - 17.96  &  OCR-II   \\
 1.15   &  8.41          &  CR$'$   &  1.15   &  6.16 - 11.05  &  OCR-II$'$ \\
 1.20   &  8.03          &  CR      &  1.20   &  8.93          &  CR        \\
 1.25   &  10.83         &  CR      &  1.25   &  10.55         &  CR        \\
\hline
\hline
   & $\mathrm{Ta} = 10$ and $\mathrm{Q} = 5$ &  &  & $\mathrm{Ta} = 10$ and $\mathrm{Q} = 10$ &  \\
 $r$ & $\langle v_3 \theta\rangle$ & Fluid patterns & $r$ & $\langle v_3 \theta\rangle$ & Fluid patterns\\
 \hline
 1.05   &  2.50 - 113.38  &  OCR-II  &   1.05   &  3.93 - 146.8  &  OCR-II  \\
 1.10   &  3.91 - 79.86   &  OCR-II  &   1.10   &  5.63 - 110.81 &  OCR-II   \\
 1.15   &  5.52 - 53.51   &  OCR-II  &   1.15   &  6.81 - 93.93  &  OCR-II  \\
 1.20   &  8.37 - 29.75   &  OCR-II  &   1.20   &  8.63 - 68.4   &  OCR-II  \\
 1.25   &  14.62          &  CR      &   1.25   &  11.32 - 46.06 &  OCR-II  \\
 \hline
 \hline
   & $\mathrm{Ta} = 20$ and $\mathrm{Q} = 1$ &  &  & $\mathrm{Ta} = 30$ and $\mathrm{Q} = 1$ &  \\
 $r$ & $\langle v_3 \theta\rangle$ & Fluid patterns & $r$ & $\langle v_3 \theta\rangle$ & Fluid patterns\\
 \hline
 1.05   &  0.87 - 134.3   &  Chaos     &   1.05   &  1.71 - 197.41 &  OCR-II(Period two)\\
 1.10   &  2.88 - 57.3    &  OCR-I     &   1.10   &  5.42 - 86.89  &  OCR-II \\
 1.15   &  5.91 - 27.09   &  OCR-II    &   1.15   &  6.75 - 37.83  &  OCR-II   \\
 1.20   &  10.41          &  CR        &   1.20   &  12.04         &  CR       \\
 1.25   &  12.55          &  CR        &   1.25   &  14.59         &  CR       \\
 \hline
 \hline
   & $\mathrm{Ta} = 40$ and $\mathrm{Q} = 1$ &  &  & $\mathrm{Ta} = 50$ and $\mathrm{Q} = 1$ &  \\
 $r$ & $\langle v_3 \theta\rangle$ & Fluid patterns & $r$ & $\langle v_3 \theta\rangle$ & Fluid patterns\\
 \hline
 1.05   &  7.47 - 185.03   &  OCR-II   &  1.05  &  0.1 - 195.29  &  Chaos \\
 1.10   &  8.59 - 113.18   &  OCR-II   &  1.10  &  5.63 - 135.65 &  OCR-II \\
 1.15   &  9.38 - 51.39    &  OCR-II   &  1.15  &  11.03 - 74.59 &  OCR-II \\
 1.20   &  13.74           &  CR       &  1.20  &  15.97         &  CR   \\
 1.25   &  16.66           &  CR       &  1.25  &  18.83         &  CR  \\
\lasthline
\end{tabular}}
\end{table*}
\end{center}

\subsection{Oscillatory instability of 2D rolls}
\begin{figure}
\begin{center}
\includegraphics[height=!, width=0.8\textwidth]{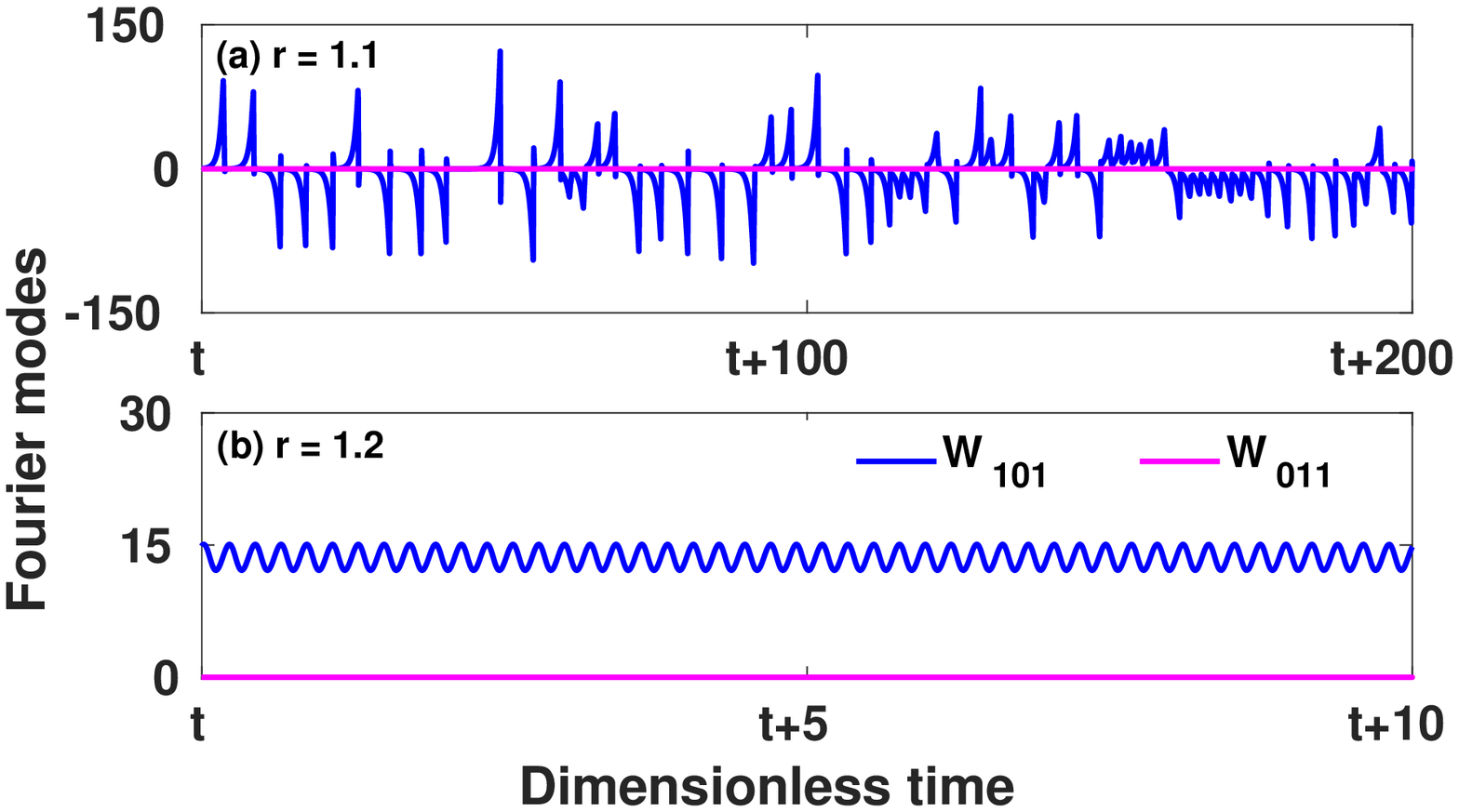}
\end{center}
\caption{Temporal evolution of the Fourier modes $W_{101}$ (blue curve) and $W_{011}$ (pink curve) computed from DNS for $\mathrm{Ta} = 5$ and $\mathrm{Q} = 1$ at two different values of $r$: (a) $r = 1.1$ and (b) $r = 1.2$.}
\label{fig:tau_5_Q_1_wavy_time_series}
\end{figure}

While performing DNS of the system using random initial conditions, we identify another class of initial conditions for which we observe chaotic as well as periodic solutions for which $W_{101}\neq 0$ but $W_{011} = 0$. The time series of these chaotic and periodic solutions are shown the figure~\ref{fig:tau_5_Q_1_wavy_time_series}. These time dependent solutions are qualitatively different than the ones mentioned in the previous subsections where both $W_{101}$ and $W_{011}$ are nonzero. To understand the origin of these solutions we perform similar analysis as done in the section~\ref{ldm} starting with the mode $W_{101}$ in vertical velocity and $Z_{010}$ in vertical vorticity. As a result, we find that seven vertical velocity modes: $W_{101}$, $W_{121}$, $W_{202}$, $W_{022}$, $W_{301}$, $W_{103}$, $W_{111}$, and twelve vertical vorticity modes: $Z_{101}$, $Z_{121}$, $Z_{202}$, $Z_{022}$, $Z_{103}$, $Z_{301}$, $Z_{010}$, $Z_{111}$, $Z_{012}$, $Z_{210}$, $Z_{020}$, $Z_{200}$ for this class of solutions contribute more than $90\%$ to the total energy. Now as before, projecting the hydrodynamic equations on these modes we get a set of $19$ nonlinear coupled ordinary differential equations.

\begin{figure}
\begin{center}
\includegraphics[height=!, width=0.8\textwidth]{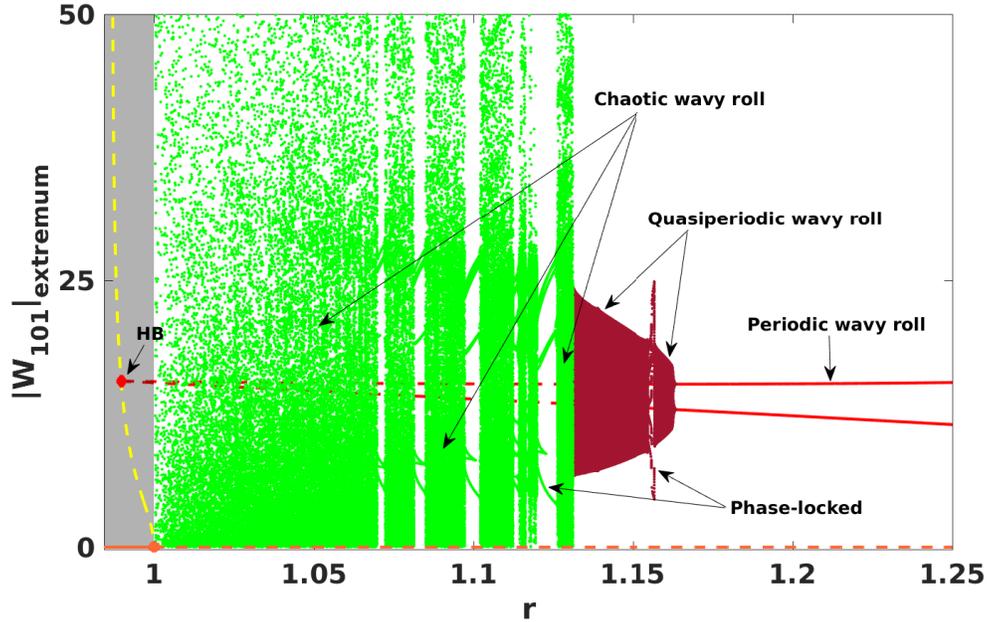}
\end{center}
\caption{Bifurcation diagram constructed using $19$-mode model for $\mathrm{Ta} = 5$ and $\mathrm{Q} = 1$ in the range $0.97\leq r \leq 1.25$. Solid and dashed curves represent the stable and unstable solutions respectively. Orange, yellow and red curves denote conduction state, steady 2D rolls along y-axis and periodic wavy rolls respectively. Brown and green dots are showing quasiperiodic and chaotic wavy rolls. A Hopf bifurcation at $r = 0.99$ on the unstable 2D rolls branch is  shown with filled red circle. An unstable limit cycle generates from there, which becomes stable via inverse Neimark-Sacker bifurcation at $r = 1.147$. As a result quasiperiodic and chaotic solutions are observed for lower values of $r$.}
\label{fig:tau_5_Q_1_wavy}
\end{figure}

\begin{figure}
\begin{center}
\includegraphics[height=!, width=0.8\textwidth]{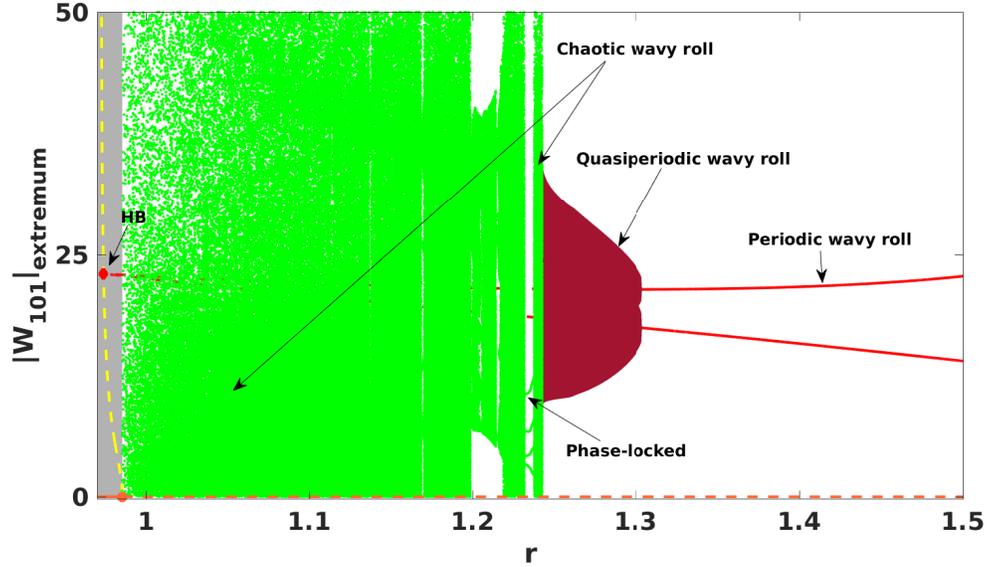}
\end{center}
\caption{Bifurcation diagram constructed for $\mathrm{Ta} = 5$ and $\mathrm{Q} = 5$. Same color coding has been used here for different solutions as it has been used in the previous one. Hopf bifurcation occurs at $r = 0.99$, whereas inverse Neimark-Sacker bifurcation takes place at $r = 1.323$ and for subsequent lower values of $r$ we see quasiperiodic and chaotic wavy rolls.}
\label{fig:tau_5_Q_5_wavy}
\end{figure}

We then perform bifurcation analysis of this $19$ dimensional model and construct two bifurcation diagrams. The bifurcation diagram constructed from this model for $\mathrm{Ta} = 5$ and $\mathrm{Q}=1$ is shown in figure~\ref{fig:tau_5_Q_1_wavy}. In this bifurcation diagram stable conduction state is shown with orange curve. Conduction state is stable (solid orange curve) for $r < 1$. At $r=1$ it becomes unstable via subcritical pitchfork bifurcation (filled orange circle in the bifurcation diagram) and an unstable 2D rolls branch is originated. The unstable conduction solutions continue to exist and those are shown with dashed orange curve in the bifurcation diagram. The unstable 2D rolls branch is shown with dashed yellow curve and it exists in the conduction regime only. This branch undergoes a Hopf bifurcation at the point marked by filled red circle in the figure~\ref{fig:tau_5_Q_1_wavy} and an unstable limit cycle is generated. The extremum values of $W_{101}$ for these unstable limit cycles have been shown with dashed red curves and they become stable via inverse Neimark-Sacker (NS) bifurcation at $r = 1.147$. The stable limit cycles are denoted by solid red curves. As a result of inverse NS bifurcation, quasiperiodic solutions are generated for lower values $r$ near $r = 1.147$. These quasiperiodic solutions are shown with brown dots in the figure~\ref{fig:tau_5_Q_1_wavy}. For further lower values of $r$ gradually chaotic solutions are obtained with intermediate phase locked states.  The chaotic solutions are shown with green dots in the bifurcation diagram. Therefore, the route to chaos at the onset is quasiperiodic in this case. These oscillatory solutions are originated via oscillatory instability of 2D rolls solutions similar to the one reported by Busse~\cite{Busse:1972} for low Prandtl-number fluids. Physically they represent wavy rolls patterns as reported in~\cite{paul:2011,dan:2015}. We then construct a bifurcation digram for $\mathrm{Ta} = 5$ and $\mathrm{Q}=5$  (see figure~\ref{fig:tau_5_Q_5_wavy}) which shows similar kind of oscillatory instability of the 2D rolls solutions. DNS results show close agreement with the model results in this case also.


\section{Conclusions}
In this paper we have investigated the bifurcation structure in zero Prandtl-number RMC by performing detailed direct numerical simulations. The range of Taylor number and Chandrasekhar number investigated in this paper are $0 < \mathrm{Ta} \leq 50$ and $0 < \mathrm{Q} \leq 50$. Interestingly we observe both time periodic and chaotic convection at the onset. The origin of  periodic as well as chaotic convecting solutions at the onset of convection is understood clearly from the bifurcation analysis of two low dimensional models constructed from the DNS data. Chaos at the onset of convection is found to occur through four different routes namely homoclinic, intermittency, and period doubling for  one set of initial conditions and  quasiperiodic for another set of initial conditions. Both type - I and III intermittency are found near the onset of convection. Chaos via intermittent and homoclinic routes are observed for weak magnetic field ($\mathrm{Q} < 2$), while period doubling route is observed for moderate magnetic field ($\mathrm{Q} \geq 2$) and larger rotation rate ($\mathrm{Ta} \sim 50$). Apart from this, periodic convection in the form of oscillatory cross rolls are also found at the onset of convection for some choices of parameter values from the considered range. The results of the model are compared with that of the DNS and we find a very good match. 

\begin{acknowledgments}
PP acknowledges support from Science and Engineering Research Board (Department of Science and Technology, India) [Grant No. EMR/2015/001680]. MG is supported by INSPIRE programme of DST, India [Code: IF150261]. Authors thank Supriyo Paul, Priyanka Maity, Yada Nandukumar, Paromita Ghosh and Ankan Banerjee for fruitful comments.
\end{acknowledgments}


\begin{thebibliography}{99}
\bibitem{marshall:1999} J. Marshall, and F. Schott, Rev. Geophys. {\bf 37}, 1–64 (1999).

\bibitem{hartman:2001} D. L. Hartmann, L. A. Moy, and Q. Fu, J. Clim. {\bf 14}, 4495–4511 (2001).

\bibitem{cardin:1994} P. Cardin, and P. Olson, Phys. Earth Planet. Inter. {\bf 82}, 235–259 (1994).

\bibitem{glatzmaier:1995} G. A. Glatzmaier, and P. H. Roberts, Nature (London) 377, 203–209 (1995).

\bibitem{cattaneo:2003} F. Cattaneo, T. Emonet, and N. Weiss, Astrophys. J. {\bf 588}, 1183–1198 (2003).

\bibitem{brent:1988} A. D. Brent, V. R. Voller, and K. J. Reid, Numer. Heat Transfer {\bf 13}, 297–318 (1988).

\bibitem{hurle:1994} D.T.J. Hurle, and R.W. Series, {\em Handbook of crystal growth}, edited by D.T.J. Hurle North Holland, Amsterdam (1994).

\bibitem{chandra:book}  {S. Chandrasekhar}, {\em Hydrodynamic and Hydromagnetic Stability}, Cambridge University Press, Cambridge (1961).

\bibitem{bodenschatz:2000} {E. Bodenschatz, W. Pesch, and G. Ahlers}, {Annu. Rev. Fluid Mech.} {\bf 32}, 709 (2000).

\bibitem{ahlers:2009} G. Ahlers, S. Grossmann, and D. Lohse, Rev. Mod. Phys., {\bf 81} 503 (2009).

\bibitem{swinney_gollub:book_1985}{F. H. Busse}, {in {\em Hydrodynamic Instabilities and the Transition to Turbulence}, Vol. {\bf 45} of Topics in Applied Physics, edited by H.L. Swinney, J.P. Gollub} Springer, Berlin, Heidelberg (1985), pp.97-133.

\bibitem{drazin:book} P. Drazin, and W. H. Reid,  {\em Hydrodynamic Stability}, Cambridge University Press, Cambridge (1981).

\bibitem{mannevile:book} P. Manneville, {\em Dissipative structures and weak turbulence}, Academic Press, San Diego (1990).

\bibitem{cross_hohenberg:1993} M. C. Cross, and  P. C. Hohenberg, Rev. Mod. Phys. {\bf 65}(3) 851-1112 (1993).

\bibitem{getling:book} A. V. Getling, {\em Rayleigh-B\'{e}nard convection: structures and dynamics} (Vol. 11), World Scientific (1998).

\bibitem{schluter:1965} A. Schl\"{u}ter, D. Lortz, and F. Busse, J. Fluid Mech. {\bf 23}(1) 129-144 (1965).

\bibitem{busse:1978} F. H. Busse, Rep. Prog. Phys. {\bf 41} 1929 (1978).

\bibitem{busse:1989} F. H.Busse, Mantle convection: plate tectonics and global dynamics, {\bf 4}, 23-95 (1989).

\bibitem{busse:1971} F. H. Busse, and J. A. Whitehead, J. Fluid Mech. {\bf 47} 305-320 (1971).

\bibitem{croquette1:Contemp.Phys_1989} {V. Croquette}, {Contemp. Phys.} {\bf 30}, 113 (1989).

\bibitem{croquette2:Contemp.Phys_1989} {V. Croquette}, {Contemp. Phys.} {\bf 30}, 153 (1989).

\bibitem{Busse:1972} {F. H. Busse}, {J. Fluid Mech.} {\bf 52}, 97 (1972).

\bibitem{clever:POF_1990} R. M. Clever, and F. H. Busse, Phys. Fluids A  {\bf 2} 334 - 339 (1990).

\bibitem{thual:1992} {O. Thual}, {J. Fluid Mech.} {\bf 240}, 229 (1992).

\bibitem{kft} {K. Kumar, S. Fauve, and O. Thual}, {J. Phys. II (France)} {\bf 6}, 945 (1996).

\bibitem{mishra:2010} {P. K. Mishra, P. Wahi, and M. K. Verma}, {Europhys. Lett.} {\bf 89}, 44003 (2010).

\bibitem{pal:2002} P. Pal, and K. Kumar, {Phys. Rev. E} {\bf 65}(4), 047302 (2002).

\bibitem{pal:2009} {P. Pal, P. Wahi, S. Paul, M. K. Verma, K. Kumar, and P. K. Mishra}, {Europhys. Lett.} {\bf 87}, 54003 (2009).

\bibitem{paul:2011} {S. Paul, P. Pal, P. Wahi, and M. K. Verma}, {Chaos} {\bf 21}, 023118 (2011).

\bibitem{pal:2013} P. Pal, K. Kumar, P. Maity, and S. K. Dana, Phys. Rev. E {\bf 87} 023001 (2013).

\bibitem{dan:2014} S. Dan, P. Pal, and K. Kumar, Eur. Phys. J. B {\bf 87} 278 (2014).

\bibitem{dan:2015} S. Dan, Y. Nandukumar, and P. Pal, Phys. Scr. {\bf 90} 035208 (2015).

\bibitem{nandu:2016} Y. Nandukumar, and P. Pal, {Computers and Fluids} {\bf 138}, 61-66 (2016).

\bibitem{spiegel:JGR_1962} {E. A. Spiegel}, {J. Geophys. Res.} {\bf 67}, 3063 (1962).

\bibitem{rossby:JFM_1969} {H. T. Rossby}, {J. Fluid Mech.} {\bf 36}, 309 (1969).

\bibitem{Roberts:1975} P.H. Roberts, and K. Stewartson, {J. Fluid Mech.} {\bf 68}, 447-466 (1975).

\bibitem{nakagawa_a:1955} Y. Nakagawa, and P. Frenzen, {Tellus} {\bf 7}(1), 1-21 (1955).

\bibitem{nakagawa:1955} Y. Nakagawa, {Nature} {\bf 175} (4453), 417-419 (1955).

\bibitem{nakagawa:1957} Y. Nakagawa, {Proc. R. Soc. Lond.} A {\bf 242}, 81-88 (1957). 

\bibitem{nakagawa:1959} Y. Nakagawa, {Proc. R. Soc. Lond.} A {\bf 249}, 138-145 (1959).

\bibitem{busse:1983} F. H. Busse, and R. M. Clever, {J. m\'{e}canique th\'{e}orique et appliqu\'{e}e} {\bf 2}, 495-502 (1983).

\bibitem{fauve_prl:1984} S. Fauve, C. Laroche, A. Libchaber, and B. Perrin, {Phys. Rev. Lett.} {\bf 52}, 1774-1777 (1984).

\bibitem{fauve:1984} S. Fauve, C. Laroche, and A. Libchaber, {J. Phys. Lett.} {\bf 45}, 101-105 (1984).

\bibitem{burr:2002} U. Burr, and U. M\"{u}ller, {J. Fluid Mech.} {\bf 453}, 345-369 (2002).

\bibitem{yanagisawa:2010} T. Yanagisawa, Y. Yamagishi, Y. Hamano, Y. Tasaka, K. Yano, J. Takahashi, and Y. Takeda, {Phys. Rev. E} {\bf 82}, 056306 (2010).

\bibitem{Eltayeb:1972} I.A. Eltayeb, {Proc. R. Soc. Lond.} A {\bf 326}, 229-254 (1972).

\bibitem{Soward:1980} A.M. Soward, {J. Fluid Mech.} {\bf 98}, 449-471 (1980).

\bibitem{Olson:2001} J.M. Aurnou, and P.L. Olson, {J. Fluid Mech.} {\bf 430}, 283-307 (2001).

\bibitem{zhang:2004} K. Zhang, M. Weeks, and P. Roberts, Phys. Fluids {\bf 16}, 2023 (2004).

\bibitem{Baig:2008} H. Varshney, and M.F. Baig, {Int. J. Heat Mass Transfer} {\bf 51}, 4095-4108 (2008).

\bibitem{Baig1:2008} H. Varshney, and M.F. Baig, {J. Turbulence} {\bf 9}, N33 (2008).

\bibitem{Podvigina:2008} O. Podvigina, {Geo. Astro. Fluid Dynamics} {\bf 104.1}, 1-28 (2008).

\bibitem{Podvigina:2010} O. Podvigina, {Phys. Rev. E} {\bf 81}, 056322 (2010).

\bibitem{Eltayeb:2013} I.A. Eltayeb, and M.M. Rahman, {Phys. of the Earth and Planetary Interiors} {\bf 221}, 38-59 (2013).

\bibitem{pal:2012} P. Pal, and K. Kumar, {Eur. Phys. J. B} {\bf 85}(6), 1-6 (2012). 

\bibitem{hirdesh:2013} H.K. Pharasi, and K. Kumar, {Phys. Fluids} {\bf 25}(10) 104105 (2013).

\bibitem{priyanka:2013} P. Maity, K. Kumar, and P. Pal, {Europhys. Lett.} {\bf 103}(6) 64003 (2013).

\bibitem{priyanka:2014} P. Maity, and K. Kumar, {Phys. Fluids} {\bf 26}(10), 104103 (2014).

\bibitem{arnab:2014} A. Basak, R.Raveendran, and K. Kumar, {Phys. Rev. E} {\bf 90}(3), 033002 (2014).

\bibitem{nandu:2015} Y. Nandukumar, and P. Pal, {Europhys. Lett.} {\bf 112.2} 24003 (2015).

\bibitem{arnab:2015} A. Basak, and K. Kumar, {Eur. Phys. J. B} {\bf 88}(10), 1-10 (2015).

\bibitem{arnab:2016} A. Basak, and K. Kumar, {Chaos} {\bf 26}(12), 123123 (2016).

%
%
%

%
%

%
%
%
%
%

\bibitem{mkv:code}{M. K. Verma, A. Chatterjee, K. S. Reddy, R. K. Yadav, S. Paul, M. Chandra, and R. Samtaney}, {Pramana} {\bf 81}, 617 (2013).

\bibitem{dhooge:matcont_2003} {A. Dhooge, W. Govaerts, and Y. A. Kuznetsov}, {ACM TOMS} {\bf 29}, 141 (2003).


%
\bibitem{veronis:1959} G. Veronis, J. Fluid Mech. {\bf 5}, 401 (1959).

\bibitem{knobloch:2013} C. Beaume, H.-C. Kao, E. Knobloch, and A. Bergeon, Phys. Fluids {\bf 25}, 124105 (2013).

\bibitem{kaplan:2017} E. J. Kaplan, N. Schaeffer, J. Vidal, and P. Cardin, Phys. Rev. Lett. {\bf 119}, 094501 (2017).

\bibitem{glendinning:2001} P. Glendinning, J. Abshagen, and T. Mullin, Phys. Rev. E, {\bf 64}(3), 036208 (2001).

\bibitem{abshagen:2001} J. Abshagen, G. Pfister, and T. Mullin, Phys. Rev. Lett. {\bf 87}(22), 224501 (2001).

\bibitem{marques:2002} F. Marques, J. M. Lopez, and V. Iranzo, Phys. Fluids {\bf 14}(6), L33-L36 (2002).

\bibitem{mkv:2017} M. K. Verma, A. Kumar, and A. Pandey, {N. J. Phys} {\bf 19}, 025012 (2017).
\end{thebibliography}
\end{document}